\documentclass[reqno, twocolumn, 10pt]{wlscirep}
\usepackage[utf8]{inputenc}
\usepackage[T1]{fontenc}
\usepackage{quotes}
\usepackage{sidecap}
\usepackage{comment}
\sidecaptionvpos{figure*}{t}

\usepackage{graphicx}
\usepackage{multirow}
\usepackage{amsmath,amssymb,amsfonts}
\usepackage{amsthm}
\usepackage{mathrsfs}
\usepackage[title]{appendix}
\usepackage{textcomp}
\usepackage{manyfoot}
\usepackage{booktabs}
\usepackage{algorithm}
\usepackage{algorithmicx}
\usepackage{algpseudocode}
\usepackage{listings}
\usepackage{placeins}
\usepackage{lmodern}
\usepackage{fix-cm}
\usepackage{color}

\usepackage{makecell}
\usepackage{subcaption}
\usepackage{bbding}
\usepackage{float}
\usepackage{hyperref}

\title{Forecasting Faculty Placement from Patterns in Co-authorship Networks}

\author[1,*]{Samantha Dies}
\author[1]{David Liu}
\author[1,2,3]{Tina Eliassi-Rad}

\affil[1]{Khoury College of Computer Sciences, Northeastern University, 440 Huntington Ave, Boston, MA 02115, USA}

\affil[2]{Network Science Institute, Northeastern University, 177 Huntington Ave., Boston, MA 02115, USA}

\affil[3]{Santa Fe Institute, 1399 Hyde Park Road, Santa Fe, NM 87501, USA}

\affil[*]{\href{mailto:dies.s@northeastern.edu}{dies.s@northeastern.edu}}

\begin{abstract}
Faculty hiring shapes the flow of ideas, resources, and opportunities in academia, influencing not only individual career trajectories but also broader patterns of institutional prestige and scientific progress. While traditional studies have found strong correlations between faculty hiring and attributes such as doctoral department prestige and publication record, they rarely assess whether these associations generalize to individual hiring outcomes, particularly for future candidates outside the original sample. Here, we consider faculty placement as an individual-level prediction task. Our data consist of temporal co-authorship networks with conventional attributes such as doctoral department prestige and bibliometric features. We observe that using the co-authorship network significantly improves predictive accuracy by up to $10\%$ over traditional indicators alone, with the largest gains observed for placements at the most elite (top-$10$) departments. Our results underscore the role that social networks, professional endorsements, and implicit advocacy play in faculty hiring beyond traditional measures of scholarly productivity and institutional prestige. By introducing a predictive framing of faculty placement and establishing the benefit of considering co-authorship networks, this work provides a new lens for understanding structural biases in academia that could inform targeted interventions aimed at increasing transparency, fairness, and equity in academic hiring practices.
\end{abstract}

\begin{document}

\flushbottom
\maketitle
\thispagestyle{empty}

\section{Introduction}
\label{sec:introduction}

Securing a tenure-track faculty position is widely regarded as a pivotal point in an early-career researcher’s trajectory. Beyond its symbolic value, such an appointment provides long-term access to institutional resources, mentorship, and intellectual visibility and serves as a mechanism of cumulative advantage. In United States (US) research institutions in particular, the prestige of a researcher’s initial faculty department can shape their career trajectory by influencing how their work is perceived, affording access to top students and collaborators, and opening doors to future opportunities~\cite{abramo.2009, zhang.2022, petersen.2015}.

However, faculty placement is not solely merit-based. While factors such as institutional prestige and prominent collaborators are often taken as proxies for merit, extensive research shows that these attributes exert effects on hiring outcomes beyond individual productivity or achievement~\cite{fernandes2019insights, cole.1974, manis.1951, hagstrom.1968, morgan.2022}. Because access to mentors and institutional resources varies systematically between candidates, hiring outcomes reflect structural inequalities as much as scholarly promise. These factors can reinforce existing hierarchies by influencing visibility, productivity, and integration into elite research communities~\cite{celis2018making, zuo.2019}. As a result, faculty hiring may perpetuate structural inequalities and contribute to stratification in academia~\cite{posselt.2020, holley.2012}.

\begin{figure*}[t!]
\centering
\includegraphics[width=\textwidth]{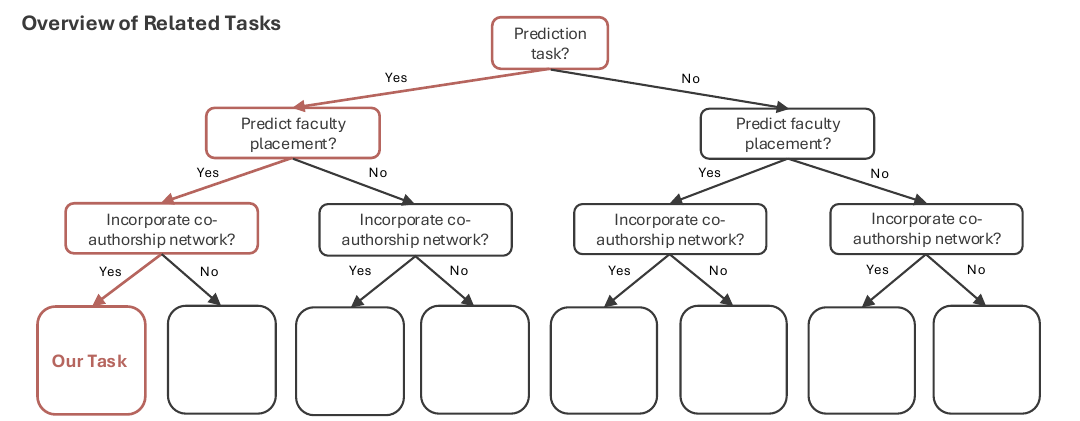}
\setlength{\unitlength}{.75cm}
    \begin{picture}(0,0)
    % No, No, No
        \put(8.9, 2.6){\makebox(0,0)[l]{\textbf{ \cite{morgan.2022}\linebreak[0]\:\cite{long.1978}\linebreak[0]\:\cite{way.2017}}}}
        
        \put(8.9, 2.2){\makebox(0,0)[l]{\textbf{\cite{abramo.2009}\linebreak[0]\:\cite{aguinis.2018}\linebreak[0]\:\cite{ceci.2015}}}}
        
        \put(8.9, 1.8){\makebox(0,0)[l]{\textbf{\cite{xing.2022}\linebreak[0]\:\cite{zhang.2023}\linebreak[0]\:\cite{laberge.2023}}}}     
        
        \put(8.9, 1.4){\makebox(0,0)[l]{\textbf{\cite{huang.2020}\linebreak[0]\:\cite{fernandes2019insights}\linebreak[0]\:\cite{pinheiro2017take}}}} 
        \put(8.9, 1.0){\makebox(0,0)[l]{\textbf{\cite{lienard2018intellectual}\linebreak[0]\:\cite{morgan.2018}\linebreak[0]\:\cite{ceci.2023}}}}

    % No, No, Yes
        \put(3.3, 2.6){\makebox(0,0)[l]{\textbf{ \cite{sun.2023}\linebreak[0]\:\cite{miller.2005}\linebreak[0]\:\cite{wapman.2022}}}}
        
        \put(3.3, 2.2){\makebox(0,0)[l]{\textbf{\cite{burris.2004}\linebreak[0]\:\cite{hargens.1967}\linebreak[0]\:\cite{bedeian.2010}}}}
        
        \put(3.3, 1.8){\makebox(0,0)[l]{\textbf{\cite{fitzgerald.2023}}}}  

    % No, Yes, No
        \put(6.1, 2.6){\makebox(0,0)[l]{\textbf{ \cite{baruffaldi2016productivity}\linebreak[0]\:\cite{yadav.2023}\linebreak[0]\:\cite{petersen.2015}}}}
        
        \put(6.1, 2.2){\makebox(0,0)[l]{\textbf{\cite{li.2022}\linebreak[0]\:\cite{sekara.2018chaperone}\linebreak[0]\:\cite{li2013co}}}}
        
        \put(6.1, 1.8){\makebox(0,0)[l]{\textbf{\cite{li2020scientific}\linebreak[0]\:\cite{hollis2001co}\linebreak[0]\:\cite{xu2020co}}}}     
        
        \put(6.1, 1.4){\makebox(0,0)[l]{\textbf{\cite{li2019early}\linebreak[0]\:\cite{collins2019hidden}\linebreak[0]\:\cite{jaramillo2023structure}}}}  

        \put(6.1, 1.0){\makebox(0,0)[l]{\textbf{\cite{jaramillo.2021}\linebreak[0]\:\cite{zuo2021understanding}\linebreak[0]\:\cite{li2009searching}}}}
    
    % No, Yes, Yes
        \put(0.5, 2.6){\makebox(0,0)[l]{\textbf{ \cite{zuo.2019}\linebreak[0]\:\cite{barnes2025edge}}}}

    % Yes, No, No
        \put(-2.3, 2.6){\makebox(0,0)[l]{\textbf{ \cite{lee2019predicting}\linebreak[0]\:\cite{way.2016}\linebreak[0]\:\cite{zhang.2022}}}}
        
        \put(-2.3, 2.2){\makebox(0,0)[l]{\textbf{\cite{daud2017finding}\linebreak[0]\:\cite{ding2018rising}\linebreak[0]\:\cite{billah2015social}}}}

    % Yes, No, Yes
        \put(-7.8, 2.6){\makebox(0,0)[l]{\textbf{ \cite{way.2016}\linebreak[0]\:\cite{clauset.2015}}}}

    % Yes, Yes, No
        \put(-5.1, 2.6){\makebox(0,0)[l]{\textbf{ \cite{sarigol2014predicting}\linebreak[0]\:\cite{grodzinski2021can}\linebreak[0]\:\cite{nikolentzos2021can}}}}
        
        \put(-5.1, 2.2){\makebox(0,0)[l]{\textbf{\cite{ductor2014social}\linebreak[0]\:\cite{du2021predicting}\linebreak[0]\:\cite{daud2017finding}}}}
        
        \put(-5.1, 1.8){\makebox(0,0)[l]{\textbf{\cite{ding2018rising}\linebreak[0]\:\cite{billah2015social}}}}
    \end{picture}
\caption{\textbf{Overview of related tasks.} We summarize the existing literature landscape according to the type of task, i.e., predictive or not. We also consider whether the models predict faculty placement and if they incorporate placement in the co-authorship network. To our knowledge, our paper is the first to frame the prediction of faculty placement using temporal co-authorship networks as a machine learning task that generalizes to unseen data.}
\label{fig:lit_taxonomy}
\end{figure*}
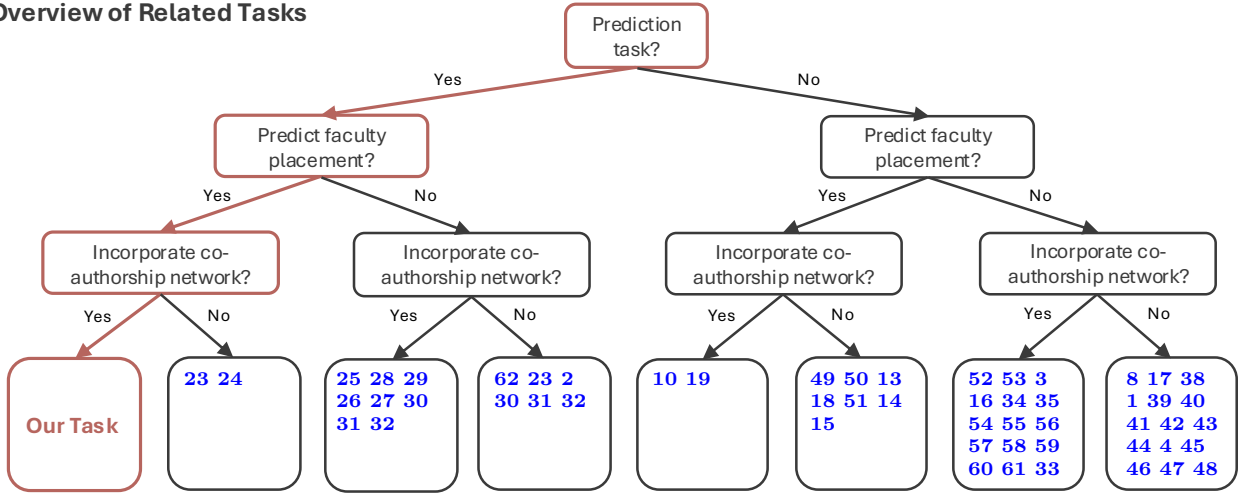

Most existing research on faculty placement is \textit{descriptive}, that is, focused on establishing associations or correlations within observed samples, rather than \textit{predictive} in the sense of assessing generalization to out-of-sample cases. These works aim to establish associations between candidate attributes (such as PhD department prestige, publication record, and network connections) and hiring outcomes~\cite{wapman.2022, zhang.2022, petersen.2015, bedeian.2010, fitzgerald.2023, li.2022, long.1978, burris.2004, barnes2025edge}. For example, Burris~\cite{burris.2004} found a strong relationship between doctoral prestige and faculty placement. Other studies, such as Long~\cite{long.1978} and Zhang et al.\cite{zhang.2022}, have investigated how productivity and citation patterns relate to hiring outcomes. Most of these studies rely on variables such as publication counts, institutional prestige, and candidate demographics, and evaluate outcomes in terms of faculty placement prestige or general research success (see Fig.~\ref{fig:lit_taxonomy}).

While these studies have greatly enhanced our understanding of academic stratification, their methods are typically correlational and retrospective: they reveal population-level trends but do not test whether these trends are strong enough to predict individual placement outcomes, especially for candidates who are not included in the data. This distinction is crucial when seeking to understand what signals hiring committees may (implicitly) rely on, and for designing interventions for fairness or transparency. Analogous efforts to move from description to prediction have transformed fields far beyond science-of-science. For example, epidemiologists increasingly deploy predictive models for individual patient risk~\cite{wynants2020prediction}, economists use forecasting to predict individual employment or credit default~\cite{athey2018impact}, and meteorologists rely on predictive models for weather and climate forecasting~\cite{bauer2015quiet}. Across these fields, prediction enables more actionable interventions and more direct tests of theory than population-level correlations alone.

A few papers have explored whether faculty placement can be predicted from pre-hire information~\cite{way.2016, clauset.2015}. For example, Clauset et al.\cite{clauset.2015} demonstrated that faculty hiring patterns closely mirror prestige hierarchies and that PhD department prestige is highly associated with placement outcomes. Another study by Way et al.~\cite{way.2016} extended this analysis to explore the predictive value of additional candidate attributes. However, these works rely on evaluation setups that are well-suited to characterizing population trends, but may present an optimistic view of true predictive performance for unseen individuals. This leaves open important questions about which pre-hire signals generalize most robustly to new candidates, and whether new features or modeling frameworks can yield deeper predictive insight. The distinctions between these descriptive and predictive tasks are summarized in Figure~\ref{fig:lit_taxonomy}.

Figure~\ref{fig:lit_taxonomy} reveals that while many descriptive studies examine the relationship between co-authorship networks and research success, only two explore how these networks relate to faculty placement. One study by Zuo et al.~\cite{zuo.2019} investigates co-authorship and placement within iSchools, while another by Barnes et al.~\cite{barnes2025edge} examines how co-authorship centrality influences the prestige of a computer science researcher’s initial faculty position. Both use regression models applied to their full set of early-career researchers, identifying correlations between network features and placement outcomes. However, neither assesses the ability of their models to generalize to researchers outside the training set. By this generalization-based definition of prediction, we are unaware of any prior work that incorporates co-authorship networks into predictive models of faculty placement.

Beyond placement, co-authorship networks have proven highly informative for forecasting downstream outcomes such as citation count~\cite{sarigol2014predicting}, research productivity~\cite{ductor2014social, du2021predicting}, and $h$-index~\cite{grodzinski2021can, nikolentzos2021can}. Some studies have even attempted to identify ``star researchers'', often defined in terms of $h$-index, citation counts, or awards, early in their careers by using co-authorship networks to compensate for limited publication history~\cite{daud2017finding, ding2018rising, billah2015social, li2009searching}. These studies show that network position, such as centrality, connectedness to high-impact researchers, or early collaborations, can be strong predictors of future research success. This raises a natural question of whether co-authorship networks also improve predictions of faculty placement.

We propose that co-authorship networks offer a new lens for investigating faculty placement. Prior work has found, for example, that having co-authors who previously published at a target venue increases one's likelihood of publishing there~\cite{sekara.2018chaperone}, and that early collaborations with highly cited researchers can ``boost'' future citation impact~\cite{petersen.2015, li.2022}. For early-career researchers, co-authorships may also signal informal advocacy and social capital, or serve as proxies for uncovering a researcher's recommendation letter writers. While these features are not directly measured in typical \textit{curriculum vitae} data, they are visible to hiring committees~\cite{li2013co, ponomariov2016co, borgatti1998network}. Hiring committees may thus evaluate a candidate’s social context alongside their individual productivity. By modeling the structure and evolution of these networks over time, we aim to approximate the network of perceived endorsements that a hiring committee might implicitly observe.

Specifically, we ask: Does a researcher’s position in the co-authorship network prior to their first hire serve as a proxy for their recommendation letter writers and help predict the prestige of their initial faculty placement? As a case study, we focus on the field of computer science (CS), which offers a large PhD pipeline, a structured placement market, and rich co-authorship data. We compare the relative predictive value of three types of pre-hire features: PhD department rank, bibliometric indicators (e.g., publication count, co-author count), and temporal co-authorship network structure. We hypothesize that, beyond the strong population-level signals from PhD rank and publication history, a candidate’s co-authorship network position provides complementary and meaningful predictive power.

\paragraph{Contributions}
Our work makes three core contributions:

\begin{enumerate}
    \item We reframe faculty placement as an individual-level prediction task, leveraging temporal co-authorship network data to forecast the prestige of a researcher’s initial appointment using only pre-hire information.
    \item We show that incorporating co-authorship features significantly improves predictive accuracy over using PhD rank alone (by $8.48\%$, $p=0.005$), bibliometric features alone (by $10.08\%$, $p<0.001$), and both combined (by $7.32\%$, $p=0.003$).
    \item We find that the predictive value of co-authorship structure is most pronounced for distinguishing placement at the most elite (top-$10$) departments, but diminishes when predicting hire at less selective definitions of high-rank placement.
\end{enumerate}

These findings support existing work which draws correlations between co-authorship network properties and faculty placement~\cite{zuo.2019, barnes2025edge}, finding that the networks encode meaningful predictive signals about how researchers are perceived and evaluated by hiring committees. More broadly, our results highlight the social and structural dimensions of faculty hiring, raising important questions about fairness and transparency in academic placement. By reframing faculty hiring as a predictive problem rooted in temporal network structure, we offer new tools for understanding academic mobility and institutional stratification. Identifying structural signals linked to placement also lays the foundation for future interventions in mentoring, networking, or hiring transparency that could democratize access to academic influence.

%
% Figure 1: Overview of related tasks
%

\section{Problem Definition: Predicting Faculty Placement with Co-authorship Networks}
\label{sec:task}

We formalize the task of predicting faculty placement as an individual-level, out-of-sample\footnote{\textit{Out-of-sample} is a machine learning term which refers to making predictions on individuals who were not included in a model’s training data, thereby assessing the model’s ability to generalize to unseen cases.} prediction problem using only information available prior to an individual being hired. Unlike descriptive analyses that examine group-level correlations, this approach asks: to what extent can one generalize to new individuals, and which features---PhD rank, bibliometric data, or co-authorship structure---carry the most predictive signal?

\subsection{Preliminaries and Mathematical Setup}
\label{sec:task:preliminaries}

We represent scholarly collaboration as a temporal co-authorship network. Let $G = \{G_{t_0}, G_{t_1}, \ldots, G_{t_f}\}$ denote a sequence of graph snapshots, with each $G_t = (\mathcal{V}, \mathcal{E}_t)$ representing the undirected co-authorship network observed up to and including year $t$. The sequence is constructed at yearly intervals, so each $G_t$ corresponds to a cumulative snapshot for a specific year. The node set $\mathcal{V}$ is fixed and consists of researchers in our cohort. The network evolves cumulatively, with new edges added as papers are published. For each researcher $v_i \in \mathcal{V}$, we construct a time-indexed feature matrix $\mathbf{X}_i \in \mathbb{R}^{d \times T}$, where $d$ is the number of features and $T$ is the number of years in the observation window. Each vector $\mathbf{x}_i(t) \in \mathbb{R}^{d}$ contains features available up to year $t$.

We design two node-level feature sets that mirror information available to hiring committees before a candidate’s first appointment. These include:
\begin{enumerate}
    \item \textit{PhD department prestige}: the prestige of the candidate's doctoral department, captured using department rank according to CSRankings~\cite{Berger.2023}, and
    \item \textit{Bibliometric features}: e.g., number of papers published, proportion of first-authored papers, average number of authors per paper, and related productivity measures.
\end{enumerate}
Full details on feature engineering, including the creation of time-dependent bibliometric features, are provided in Section~\ref{sec:methodology:data}. Crucially, we consider bibliometric features to be ones which we can access from a paper by author bipartite network. This means the bibliometric features include some co-authorship information, such as average number of co-authors per paper and number of papers with at least one faculty co-author. On the other hand, our co-authorship network features capture \textit{placement} in the network.

We define each researcher’s outcome variable, \textit{faculty placement}, as the prestige $y_i \in \{\texttt{High}, \texttt{Not High}\}$ of their first tenure-track appointment. We use CSRankings~\cite{Berger.2023}, a publication-based metric specific to Computer Science departments within the US, to assign numerical ranks that serve as proxies for institutional prestige. The publication-based rankings align naturally with our task, as co-authorship networks are also derived from publication data. Moreover, we avoid other measures of institutional prestige, such as those derived from faculty hiring networks, because of potential circularity between the labels we predict in our task and the network features used in our models. We discretize these ranks to define high-rank placements (e.g., top-$10$, top-$20$). While prestige is a continuous construct, prior work has shown that top tier departments exhibit qualitatively different hiring patterns\cite{clauset.2015, way.2016, burris.2004}. As such, this binary framing reflects both practical concerns (class balance, stability) and theoretical motivation (distinctive hiring dynamics at the top tier).

\subsection{Prediction Task Definition}
\label{sec:results:task_definition}
Let $\mathcal{V}_\mathrm{hire} \subset \mathcal{V}$ denote the set of researchers who receive their first faculty placement during the observation window (i.e., between $t_0$ and $t_f$). For each $v_i \in \mathcal{V}_\mathrm{hire}$ hired in year $t_i$, the goal is to predict $y_i$ using only data available prior to their hire. Formally,
$$\hat{y}_i = f \left(\{\mathbf{x}_i(t_0),\dots,\mathbf{x}_i(t_i-1)\}, \{G_{t_0},\dots,G_{t_i-1}\} \right),$$
where $f$ is a machine learning model trained to generalize to unseen individuals. The task assumes that each researcher's hire year $t_i$ is known and focuses on predicting \textit{where} they are hired, rather than \textit{whether} they are hired at all.

This setup enables us to test two core questions: (1) Is faculty placement, as an individual outcome, predictable from pre-hire data? (2) If so, which features---PhD rank, bibliometric information, co-authorship structure, or some combination of the three---are most informative?

%
% Figure 2: Temporal data splits and the modeling pipeline
%
\begin{figure*}[p!]
\centering
\includegraphics[width=\textwidth]{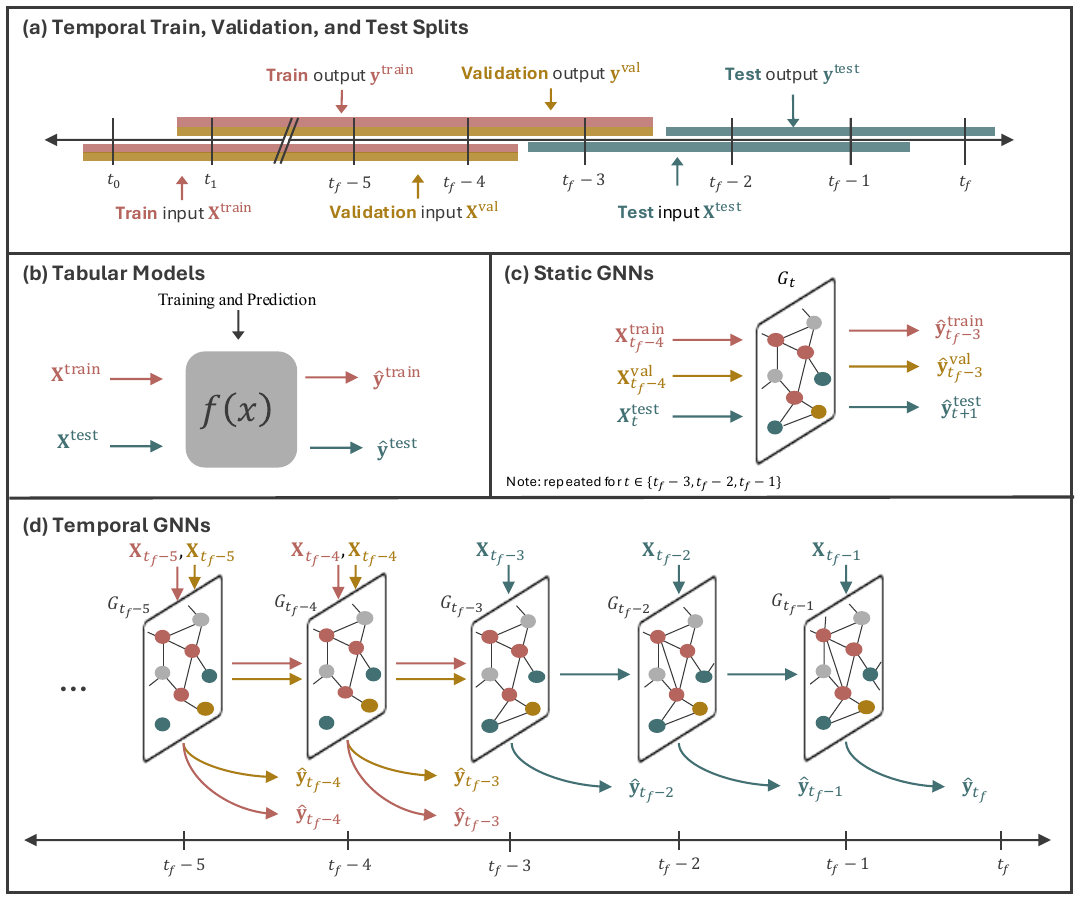}
\caption{\textbf{Temporal data splits and the modeling pipeline.} We depict how we split the data into train, validation, and test sets and outline how the datasets are used by different models. Here, \textbf{(a)} displays the temporal data splitting process. Data from years $t_0$ to $t_f-4$ are used as the train (red) and validation (yellow) input, while data from years $t_f-3$ to $t_f-1$ serve as the test (blue) input. In all cases, we define $\mathbf{y}$, the class label in $\{\texttt{High},\,\texttt{Not High}\}$, using data from the year after the corresponding input (e.g., $\mathbf{X}_{t_0}$ is evaluated with $\mathbf{y}_{t_1}$). As a result, the test input for year $t_f - 3$ overlaps in time with the training and validation output. However, this does not constitute data leakage, as the class labels are derived solely from observed placement outcomes and do not depend on any of the input features used for prediction. In \textbf{(b)} tabular models, e.g., Random Forest, the full training data $\mathbf{X}^\mathrm{train}$ is fed in at once and $\mathbf{X}^\mathrm{test}$ is used at testing time to generate $\mathbf{\hat{y}}^\mathrm{test}$. For the \textbf{(c)} static GNNs, we train three separate models on the co-authorship network snapshots $G_{t_f-3}$, $G_{t_f-2}$, and $G_{t_f-1}$. In all cases, we use $\mathbf{X}^\mathrm{train}_{t_f-4}$ to train and $\mathbf{X}^\mathrm{val}_{t_f-4}$ to validate. At the testing stage, we use the test set for the year corresponding to the network year. Finally, for the \textbf{(d)} temporal GNNs, we train on the co-authorship snapshots $\{G_{t_0},\dots,G_{t_f-4}\}$ with the corresponding train and validation data. The model weights are updated using \textit{backpropagation through time}~\cite{werbos.1990}. We then produce the predicted class labels $\mathbf{\hat{y}}^\mathrm{test}$ by passing the model over $G_{t_f-3}$, $G_{t_f-2}$, and $G_{t_f-1}$ with the corresponding test input. Our pipeline protects against temporal data leakage and maximizes consistency across differing model architectures.}
\label{fig:methods}
\end{figure*}

\section{Methodology}
\label{sec:methodology}

This section describes how we construct the dataset, define input features, and formalize the prediction task. We first detail how we combine publication records with faculty hiring metadata to build temporal co-authorship networks and extract pre-hire features (Section~\ref{sec:methodology:data}). We then describe our modeling framework, including tabular classifiers, graph neural networks, and a spatiotemporal GNN, along with the training and evaluation strategy used to ensure a fair comparison (Section~\ref{sec:methodology:modeling}).

\subsection{Data}
\label{sec:methodology:data}

We construct a dataset by merging publication records, faculty metadata, and institutional rankings. This section describes how we preprocess and combine these sources to generate the co-authorship network, faculty placement labels, and node-level features used in our models. We also describe how we split the data into train, validation, and test sets for our machine learning task in a way which prevents temporal data leakage.

\subsubsection{Data Aggregation}

We construct our dataset by integrating three sources: (1) DBLP~\cite{dblp}, publication records for $1.8$ million publications by $1.3$ million authors from $1947$–$2023$, (2) the CS Professors dataset~\cite{Huang.2022}, which includes names, universities, hire years, subfields, and PhD institutions for $5323$ US computer science faculty, and (3) CSRankings~\cite{Berger.2023}, US computer science department prestige ranks based on publication data.

We focus on researchers who appear in the CS Professors dataset---documented tenure-track faculty at US institutions---to center our analysis on placement prestige within academia. Faculty lacking known hire years are excluded. We then extract each researcher's publication history and co-authorship relationships from DBLP, and merge department prestige information from CSRankings to label institutional placements.

This aggregation yields a final dataset of $4656$ US-based computer science faculty members, each with complete publication histories, known tenure or tenure-track placements, and institutional prestige labels. Details of data cleaning, record linkage, and field harmonization are provided in Section~\ref{sec:SI:data_processing}.

\subsubsection{Network Construction}
\label{sec:methodology:network}

We construct a temporal co-authorship network using faculty metadata, publication data, and institutional rankings. We define the network as a sequence of cumulative annual snapshots, $G = \{G_{2010}, G_{2011}, \dots G_{2020}\}$, where each $G_t=(\mathcal{V}, \mathcal{E}_t)$ is an undirected graph representing co-authorships observed up to and including year $t$. The node set $\mathcal{V}$ is fixed and consists of the $4656$-researcher cohort from the DBLP and CS Professors data (i.e., $|\mathcal{V}| = 4656$). Edges are weighted by the number of co-authored papers: for nodes $v_i, v_j \in \mathcal{V}$, the edge weight $w_{ij}(t)$ equals the number of joint publications observed up to year $t$.

To structure our prediction task, we partition the node set into two disjoint subsets: researchers whose first tenure-track appointment occurred between $2010$ and $2020$, denoted $\mathcal{V}_\mathrm{hire}$, and researchers hired prior to $2010$, denoted $\mathcal{V}_\mathrm{faculty}$. The set $\mathcal{V}_\mathrm{hire}$ defines the pool of target nodes for prediction. Out of the $4656$ total nodes, $1974$ are hired between $2010$ and $2020$ (i.e., $|\mathcal{V}_\mathrm{hire}| = 1974$), and the remaining $2682$ nodes are established faculty members (i.e., $|\mathcal{V}_\mathrm{faculty}| = 2682$).

Although we only focus on the nodes in $\mathcal{V}_\mathrm{hire}$ in our prediction task, we generate node labels and features for all $4656$ researchers. This design serves two purposes. First, some of our features depend on the status of their co-authors in the year of publication, and this set of co-authors may include members of $\mathcal{V}_\mathrm{faculty}$. Second, in graph machine learning models, all nodes are included in the co-authorship network to preserve the structural context and enable message passing. Details on the node labels and features are listed below.

\paragraph{Node Labels}
To generate node labels, we consider the ranks of each node's faculty department according to CSRankings. We then assign a binary label $y_i \in \{\texttt{High},\,\texttt{Not High}\}$ to each researcher $v_i \in \mathcal{V}$ based on whether their first faculty appointment was at a high-rank department. Throughout this paper, we primarily define high-rank departments as those ranked in the top-$10$ (approximately $5\%$ of the $186$ departments in our data), but we also evaluate robustness to alternative thresholds, top-$20$, $30$, $40$, and $50$ (approximately $11\%$, $16\%$, $22\%$, and $27\%$ of the departments in our data, respectively), in Section~\ref{sec:results:robustness}. We choose these thresholds to reflect the qualitative differences in hiring patterns documented in previous work on prestige hierarchies~\cite{clauset.2015, burris.2004}. While labels are assigned to all researchers in $\mathcal{V}$, we restrict training, validation, and testing to those in $\mathcal{V}_\mathrm{hire}$, whose first placements occur between $2010$ and $2020$. When we define high-rank as departments in the top-$10$, $1055$ of the $4656$ researchers in $\mathcal{V}$ are hired at high-rank departments. Further, $445$ of the $1974$ researchers in $\mathcal{V}_\mathrm{hire}$ were hired at high-rank departments, leading to an overall class imbalance of $22.2\%$ researchers with $y_i = \texttt{High}$ and $77.8\%$ of researchers with $y_i = \texttt{Not High}$. As we gradually decrease the threshold for what we consider a high-rank department, we improve and reverse the class imbalance, resulting in the most equal class distribution if we define high-rank as top-$30$ ($49.1\%$ high, $50.09\%$ not high), and a final class imbalance of $77.1\%$ high and $22.9\%$ not high when high-rank refers to top-$50$ departments.

\paragraph{Node Features}
For each researcher $v_i \in \mathcal{V}$, we construct two sets of node features: (1) a static feature for PhD department rank, and (2) a set of time-dependent bibliometric features derived from their publication history. All features are defined such that only information available prior to the year of hire is used for prediction, ensuring no data leakage.
\begin{itemize}
    \item \textit{PhD Department Rank:} A static feature encoding the rank of each researcher’s PhD-granting institution (from CSRankings). This value is broadcast across the $11$ years in our sample, i.e., from $2010$ to $2020$, resulting in a feature tensor $\mathbf{X}_\mathrm{PhD} \in \mathbb{R}^{|\mathcal{V}| \times 1 \times 11}$.
    \item \textit{Bibliometric Features:} For each researcher and each year $t$, we compute a set of features characterizing their publication record, including counts, authorship positions, and select co-author attributes (such as the number of co-authors or whether any co-author is affiliated with a high-rank department as of year $t$). The bibliometric features form the tensor $\mathbf{X}_\mathrm{Bib} \in \mathbb{R}^{|\mathcal{V}| \times 22 \times 11}$, combining cumulative and prior-year values for each feature to yield a total of $22$ features across $11$ years.

    Importantly, these features incorporate attributes of co-authors, i.e., information available from the paper-author incidence matrix, but do not use any measures of a researcher’s global position within the co-authorship network (e.g., centrality, clustering coefficient). Network position features, which require the author-author adjacency matrix, are reserved for our graph-based models.
\end{itemize}

Full details on the feature definitions and additional steps to avoid data leakage are provided in Section~\ref{sec:SI:features}. These features form the basis for our prediction models. For traditional tabular learning, we use $\mathbf{X}_\mathrm{PhD}$, $\mathbf{X}_\mathrm{Bib}$, and their concatenations as input to standard classifiers such as Logistic Regression, Random Forest, and Multi-layer Perceptron. These models operate on a per-node basis and do not incorporate network structure directly. In contrast, our graph-based models take the co-authorship network $G$ as input and use message passing to aggregate information from each node's neighbors. To fairly compare across architectures, we evaluate all models using the same prediction task and define node masks to ensure that training, validation, and testing are performed only on researchers in $\mathcal{V}_\mathrm{hire}$. Details on model architectures, training splits, and implementation strategies are provided in Sections~\ref{sec:methodology:splits} and~\ref{sec:methodology:modeling}.

\subsubsection{Training, Validation, and Testing Data Splits}
\label{sec:methodology:splits}

Because our task involves predicting a researcher's initial faculty placement using only pre-hire information, we must prevent future information from leaking into our training procedure. This constraint is especially important for temporal GNNs, which operate on sequences of graph snapshots and accumulate node representations over time. To enforce a clean separation between observed and unobserved data, we define a strict, temporal cut between training and evaluation, and ensure that training and validation sets are disjoint. Figure~\ref{fig:methods}(a) depicts this temporal cut visually.

To create the train and test temporal split, we let $\mathcal{V}_\mathrm{hire}^{(y)} \subseteq \mathcal{V}_\mathrm{hire}$ denote the set of researchers hired in year $y \in \{2010, \dots, 2020\}$. We define the \textit{test set} to be all nodes hired between $2018$ and $2020$:
\begin{equation}
    \mathcal{V}_\mathrm{test} = \bigcup_{y=2018}^{2020}\mathcal{V}_\mathrm{hire}^{(y)}.
\end{equation}
The nodes hired between $2010$ and $2017$ are candidates for the train set. We define this training candidate pool as
\begin{equation}
    \mathcal{V}_\mathrm{train\_pool} = \bigcup_{y=2010}^{2017}\mathcal{V}_\mathrm{hire}^{(y)},
\end{equation}
where $\mathcal{V}_\mathrm{train\_pool}^{(y)} \subseteq \mathcal{V}_\mathrm{train\_pool}$ is the set of candidate training nodes hired in each year $y \in \{2010, \dots, 2017\}$. For each year $y \in \{2010, \dots, 2017\}$, we define $\mathcal{V}_\mathrm{train}^{(y)} \subset \mathcal{V}_\mathrm{train\_pool}^{(y)}$, sampled nodes uniformly at random with probability $p=0.8$. The remaining nodes are used for validation. Thus, our full \textit{train set} is defined as
\begin{equation}
    \mathcal{V}_\mathrm{train} = \bigcup_{y=2010}^{2017}\mathcal{V}_\mathrm{train}^{(y)}.
\end{equation}
Finally, we define our \textit{validation set} to be $\mathcal{V}_\mathrm{val} = \mathcal{V}_\mathrm{train\_pool} \setminus \mathcal{V}_\mathrm{train}$, meaning that $\mathcal{V}_\mathrm{val}^{(y)} \subseteq \mathcal{V}_\mathrm{val}$ for each year $y \in \{2010, \dots, 2017\}$. The train, validation, and test set are disjoint, and the node sets together incorporate all of the nodes hired between $2010$ and $2020$  (i.e., $\mathcal{V}_\mathrm{train} \cup \mathcal{V}_\mathrm{val} \cup \mathcal{V}_\mathrm{test} = \mathcal{V}_\mathrm{hire}$). This protocol ensures a strict temporal separation between observed (training/validation) and unobserved (test) researchers, enabling robust evaluation of predictive generalization to truly out-of-sample cases.

\subsection{Modeling}
\label{sec:methodology:modeling}

This section details our comprehensive modeling framework for predicting faculty placement. We describe the suite of models used, ranging from simple heuristics and standard machine learning classifiers to graph neural networks that incorporate the structure and temporal evolution of the co-authorship network. We further outline our strategies for feature isolation, network rewiring, and consistent training and evaluation pipelines to ensure robust and comparable assessment across all models.

%
% Table 1: Model compatibility with feature sets
%
\begin{table*}[tb]
\centering
\resizebox{\textwidth}{!}{%
\renewcommand{\arraystretch}{1.5}
\begin{tabular}{@{} l | c | c | c | c | c | c | c @{}}
\textbf{Model} & \textbf{PhD} & \textbf{Bib} & \textbf{Co-author} & \textbf{PhD + Bib} & \textbf{PhD + Co-author} & \textbf{Bib + Co-author} & \textbf{PhD + Bib + Co-author} \\
\hline \hline
Random           &  &  &  &  &  &  &  \\
PhD rank        & \checkmark &        &        &        &        &        &        \\
Avg. co-author rank       &        &        & \checkmark &        &        &        &        \\
\hline
White-box Tabular Models~\cite{cortes1995support, breiman2001random, chen2016xgboost} & \checkmark & \checkmark &  & \checkmark & &  &  \\
Tabular Neural Networks~\cite{rumelhart1986learning, huang2020tabtransformer} &  & \checkmark &  & \checkmark &        &        &        \\
Static GNNs~\cite{kipf2016semi, velickovic2017graph, hamilton2017inductive}             &     &     & \checkmark &    & \checkmark & \checkmark & \checkmark \\
Temporal GNNs~\cite{seo.2018}  &     &     & \checkmark &     & \checkmark & \checkmark & \checkmark \\
\end{tabular}
}
\caption{\textbf{Model compatibility with feature sets.} We list the types of models used in our analysis and the feature sets that they are compatible with. We compare our models against three heuristics: random guessing, PhD rank, where we assign a label of high if the PhD rank is high, and average co-author rank, where we assign a label of high if the average faculty department rank of the node's neighbors is high. The white-box tabular models, Logistic Regression, Support Vector Machines~\cite{cortes1995support}, Random Forest~\cite{breiman2001random}, and XGBoost~\cite{chen2016xgboost}, cannot directly use the co-authorship network as input. The tabular neural network models, the Multi-layer Perceptron~\cite{rumelhart1986learning} and the Transformer\cite{huang2020tabtransformer}, are designed to work with a large number of tabular features, so we train them with either bibliometric features or both PhD rank and bibliometric data. To take full advantage of the structure of the co-authorship network, we consider four graph machine learning models, GCN~\cite{kipf2016semi}, GraphSAGE~\cite{hamilton2017inductive}, GAT~\cite{velickovic2017graph}, and GConvGRU~\cite{seo.2018}. We selected a diverse set of models that span a range of inductive biases, ensuring that each model type is well suited to the structure and modality of the input data.}
\label {tab:models}
\end{table*}

\subsubsection{Model Overview}

We benchmark a diverse set of models to evaluate the predictive value of different feature sets and modeling approaches. These include simple heuristics, standard machine learning models using tabular features, and graph neural networks that incorporate co-authorship network structure (see Table~\ref{tab:models}). Below, we describe each model class and its role in our comparative analysis.

\paragraph{Heuristics}

We consider three heuristics to help contextualize the performance of our models. First, we consider how well we would perform with random guessing. Second, we consider how well we can predict faculty placement directly from PhD department rank. For this heuristic, we predict $\hat{y}_i = \texttt{High}$ if node $v_i$ has a high PhD department rank. For example, if we consider faculty departments in the top-$10$ to be high-rank, then we predict high if node $v_i$ has a PhD department rank in the top-$10$. Our third heuristic looks at the co-authors of node $v_i$ in the year before $v_i$ is hired and calculates the average faculty department rank for all co-authors that are already faculty. If the average co-author rank is considered high-rank, we predict high-rank for node $v_i$. 

We compute all three heuristics for all nodes in $\mathcal{V}_\mathrm{test}$, thereby allowing for direct comparison with the graph-based and tabular models.

\paragraph{Tabular Models}

Our first class of machine learning models treats the prediction task as a standard supervised classification problem over tabular features, without access to the temporal co-authorship network. Like with our graph-based task, each model is trained to predict whether a researcher's first faculty appointment is at a high-rank institution from data available prior to their hire year. However, unlike with the graph task, these models only use PhD prestige features $\mathbf{X}_\mathrm{PhD}$, bibliometric features $\mathbf{X}_\mathrm{Bib}$, or their concatenation. These features are drawn from the same tensors used by the GNNs, but are extracted only from the final pre-hire year for each researcher (see Fig.~\ref{fig:methods}(b)).

We benchmark a range of machine learning models with different inductive biases: Logistic Regression (\texttt{LR}) as a linear classifier, Support Vector Machines~\cite{cortes1995support} (\texttt{SVM}) for kernel-based classification, Random Forest~\cite{breiman2001random} (\texttt{RF}) and XGBoost~\cite{chen2016xgboost} (\texttt{XGB}) as tree-based ensemble methods, and a Multi-layer Perceptron~\cite{rumelhart1986learning} (\texttt{MLP}) and a tabular Transformer model~\cite{huang2020tabtransformer} (\texttt{Trans}) for neural architectures.

\paragraph{Graph-Based Models}

In addition to the tabular models, we evaluate four graph neural network architectures: Graph Convolutional Network~\cite{kipf2016semi} (\texttt{GCN}), a transductive model that applies spectral convolutions over a single static graph $G_t$; Graph Attention Network~\cite{velickovic2017graph} (\texttt{GAT}), a model similar to \texttt{GCN} that uses attention mechanisms to weight neighbor contributions; GraphSAGE~\cite{hamilton2017inductive} (\texttt{GraphSAGE}), a model that aggregates sampled neighbor features for inductive representation learning; and Graph Convolutional Gated Recurrent Unit~\cite{seo.2018} (\texttt{GConvGRU}), a spaciotemporal GNN that embeds each $G_t$ using Chebyshev graph convolutions~\cite{defferrard.2016} and feeds node embeddings into a Gated Recurrent Unit (GRU) across time.

The static models (\texttt{GCN}, \texttt{GAT}, \texttt{GraphSAGE}) operate on a single graph $G_{t_i-1}$ (Fig.~\ref{fig:methods}(c)), while \texttt{GConvGRU} is trained over temporal sequences $\{G_{t_i-w-1}, \dots, G_{t_i-1}\}$, where $w$ is the sliding window size (Fig.~\ref{fig:methods}(d)). Each model receives node features from $\mathbf{X}_\mathrm{PhD}$, $\mathbf{X}_\mathrm{Bib}$, both, or, in the case of no features, the constant vector $\mathbf{1}$ as is standard in graph machine learning~\cite{cui2022positional}. Message passing is performed over the full node set $\mathcal{V}$, but predictions and loss computation are restricted to $\mathcal{V}_\mathrm{hire}$ using binary node masks.

\subsubsection{Modeling Pipelines}

To ensure fair and rigorous evaluation, we apply a unified training and evaluation protocol across all model classes, with strict safeguards against temporal data leakage. All models use identical training, validation, and test splits based on the year of faculty appointment, ensuring that only pre-hire information is used for prediction (see Fig.~\ref{fig:methods}). Below, we describe the pipeline for each model class, progressing from tabular models to static and temporal graph neural networks.

\paragraph{Tabular Models}

For tabular models, each researcher $v_i$ hired in year $t_i$ is represented by a feature vector extracted from data available up to year $t_i-1$ (i.e., prior to their hire). Depending on the experiment, this feature vector includes PhD rank, bibliometric features, or both. Models are trained to predict whether a candidate’s first faculty appointment is at a high-rank department, using the train, validation, and test node splits described in Section~\ref{sec:methodology:splits}. Neural models are trained using cross-entropy loss, while non-neural models are tuned via grid search and cross-validation. In all cases, evaluation is conducted on a temporally held-out test set, and only information available prior to hire is used as input (see Fig.~\ref{fig:methods}(b)).

\paragraph{Static Graph Neural Networks (\texttt{GCN}, \texttt{GAT}, \texttt{GraphSAGE})}

For static GNNs, we treat each test year $t \in \{2018, 2019, 2020\}$ as a separate prediction task. Models are trained and evaluated on the co-authorship network snapshot $G_{t-1}$, using node features from the corresponding pre-hire year (see Fig.~\ref{fig:methods}(c)). As with the tabular models, features may include PhD rank, bibliometric features, or both, and all node sets (train, validation, test) are identical to those used for tabular models. Message passing is performed over the full network, but loss and evaluation are restricted to nodes hired in the relevant year. Early stopping is based on validation loss, and final performance is computed on the held-out test nodes.

\paragraph{Temporal Graph Neural Network (\texttt{GConvGRU})}

For the temporal GNN, \texttt{GConvGRU}, we implement a sliding-window approach to capture the evolving structure of the co-authorship network (see Fig.~\ref{fig:methods}(d)). The model receives sequences of $w$ consecutive network snapshots and corresponding node features, and is trained to predict outcomes for the final year in each sequence. For each training year $t$, the input consists of the network snapshots and features from years $t-w$ to $t-1$, and supervision is restricted to nodes hired in those years. During each training epoch, the model unrolls over all sequences, using a recurrent hidden state to propagate temporal information. Validation and testing follow the same masking and evaluation procedures as in the static case, ensuring consistency across model classes.

In all models, nodes hired prior to $2010$ and non-candidate nodes participate in message passing but are excluded from loss and evaluation. This unified pipeline ensures that all predictive models are trained and evaluated under consistent conditions, enabling robust comparison of feature sets and architectures.

Full details of node mask construction, sliding window logic, and optimization procedures are provided in Section~\ref{sec:SI:pipeline}. Details on the experimental setup, including hyperparameter grids and model architectures, can be found in Section~\ref{sec:SI:experiments}.

\subsubsection{Model Evaluation}

We evaluate model performance using precision, recall, and area under the precision-recall curve (PR-AUC). Precision measures the proportion of predicted \texttt{High} placements that are correct, while recall captures the proportion of actual \texttt{High} placements that are successfully identified. PR-AUC summarizes the trade-off between these two metrics across different classification thresholds using the predicted probabilities. This metric is particularly well suited for our task, which involves imbalanced class labels, as it accounts for both sensitivity and specificity without being dominated by the majority class. All evaluation metrics are computed on $\mathcal{V}_\mathrm{test}$ for each model.

To assess the contribution of each feature set while accounting for variability across model architectures, we use a linear mixed-effects regression. This approach allows us to estimate the effect of different feature sets on model performance while controlling for the fact that some model types (e.g., GCN, SVM, etc.) may systematically perform better or worse than others. The regression predicts the PR-AUC of a model–feature set combination using two components, a \textit{fixed effect} for the feature set (what we are testing), and a \textit{random effect} for the model architecture (what we want to control for but not test directly). 

We define the model formally as:
\begin{equation}
    \texttt{PR-AUC}_{ij} = \mu + \sum_{k \not = \texttt{ref}}\beta_k \cdot \mathbf{1}[\texttt{feat}_i = k] + u_j + \varepsilon_{ij},
\end{equation}
where $\texttt{PR-AUC}_{ij}$ is the performance of the $j$th model trained with the $i$th feature set; $\mu$ is the global intercept, equal to the mean PR-AUC of the reference feature set; $\beta_k$ is the fixed effect, the estimated improvement (or drop) in performance when using feature set $k$ instead of the reference; $\mathbf{1}[\texttt{feat}_i = k]$ is an indicator function that equals 1 when the feature set is $k$; $u_j \sim \mathcal{N}(0, \sigma_u^2)$ is a random intercept for each model architecture, capturing model-specific performance differences; and $\varepsilon_{ij} \sim \mathcal{N}(0, \sigma^2)$ is the residual error term.

This mixed-effects structure is essential because model architecture introduces structured, non-random variation: GCNs and MLPs, for example, differ in inductive biases and expressive power. If we ignored this by using a standard linear regression, our estimates of feature set effects would be confounded by model type. Mixed-effects models let us pool information across model types while controlling for these architectural differences, improving statistical power and avoiding misleading conclusions.

We repeat this analysis with four reference feature sets: PhD rank, the bibliometric features, PhD rank and bibliometric features, and PhD rank and the co-authorship network. This allows us isolate the added predictive value of co-authorship structure in each setting by directly comparing, for example, PhD rank \textit{and} the co-authorship network against just PhD rank. All models included in this analysis are trained using identical node splits and evaluation criteria to enable a fair comparison across the full range of feature sets.

\section{Experiments}

We now describe the experimental setup used to evaluate the predictive performance of the faculty placement models. This section covers the construction of our dataset, the engineering of the node-level feature sets, and the training procedures for both the tabular and graph-based models. All experiments are conducted using consistent temporal train, validation, and test splits, and the models are evaluated under a strict pre-hire constraint to ensure fair and leakage-free comparisons across feature sets and model classes.

\subsection{Data Processing and Aggregation}
\label{sec:SI:data_processing}

To construct the combined dataset used in our analysis, we begin by preprocessing the DBLP publication records to retain only papers that list at least one author appearing in the CS Professors dataset. This filtering step reduces noise and ensures that our publication data is relevant to our target population of US computer science faculty.

We then merge the author identities across datasets by matching the \texttt{author} field in DBLP to the \texttt{FullName} field in the CS Professors dataset. To address inconsistencies in naming conventions (e.g., initials, middle names, or ordering), we standardize names using established text processing techniques. Ambiguous or duplicate entries are manually inspected and corrected as needed to maximize match accuracy.

Next, we manually standardize university and institution names across all data sources to resolve variations (e.g., ``MIT'' vs. ``Massachusetts Institute of Technology''). For institutions not explicitly listed in the rankings, we imputed prestige scores using the average rank of researchers in the dataset.

For each faculty member, we extract all available publication records, including publication years, the full set of co-authors, and the authorship order (e.g., first author, second author, etc.). This allows us to generate bibliometric features and construct the temporal co-authorship network as described in Sections~\ref{sec:methodology:network} and~\ref{sec:SI:features}.

We exclude faculty members who lack known hire years, since our analysis framework assumes the timing of each researcher’s transition into a faculty position is observed. We also removed duplicate or incomplete records to ensure data quality.

The final dataset was formed by merging bibliometric features, co-authorship network information, faculty data, and institutional prestige labels for each researcher. All data processing steps are performed using Python. The code used to perform this data processing is available at \url{https://github.com/samanthadies/predicting-faculty-placement}.

\subsection{Node Feature Construction}
\label{sec:SI:features}

\paragraph{PhD Department Rank}
For each researcher $v_i$, we assign a static feature corresponding to the rank of their PhD-granting institution according to CSRankings. This scalar value is broadcast across all eleven years, forming the tensor $\mathbf{X}_\mathrm{PhD} \in \mathbb{R}^{|\mathcal{V}| \times 1 \times 11}$, where each slice along the time axis contains the same value.

\paragraph{Bibliometric Features}
For each researcher $v_i$ and each year $t \in \{2010, \dots, 2020\}$, we construct $11$ bibliometric features derived from DBLP publication records and the CS Professors dataset. Each feature is computed both cumulatively (up to year $t$) and for the previous year ($t-1$), yielding a tensor $\mathbf{X}_\mathrm{Bib} \in \mathbb{R}^{|\mathcal{V}| \times 22 \times 11}$. The bibliometric features are:
\begin{enumerate}
    \item Number of papers published,
    \item Average number of authors per paper,
    \item Number of first-authored papers,
    \item Proportion of first-authored papers,
    \item Average author position,
    \item Number of papers co-authored with at least one faculty member,
    \item Proportion of papers co-authored with at least one faculty member,
    \item Number of papers co-authored with a faculty member hired at a top-$10$ department,
    \item Proportion of papers co-authored with a faculty member hired at a top-$10$ department,
    \item Number of papers co-authored with a faculty member hired at a top-$50$ department, and
    \item Proportion of papers co-authored with a faculty member hired at a top-$50$ department.
\end{enumerate}
Features $2$ and $6$-$11$ use attributes of a researcher's co-authors (such as hire year or department rank), determined solely by information available up to year $t$. For example, if researcher $A$ co-authors a paper with researcher $B$ in $2012$, but $B$ is not hired to a high-rank department until $2013$, then $B$ does not count as a high-rank co-author in $2012$, but does in subsequent years.

Crucially, all features are extracted from the paper–author incidence matrix, and do not capture network position (e.g., centrality, clustering coefficient). Such network topology features are included only in models that explicitly operate on the author–author co-authorship network $G$.

\subsection{Model Training}
\label{sec:SI:pipeline}

\paragraph{Node Masks}

To ensure strict separation of training, validation, and testing---and to prevent data leakage---we construct binary node masks for each year $t$ and model split. These masks indicate which nodes are included in loss computation and evaluation at each time point and are used consistently across all model classes (tabular, static GNN, temporal GNN). 

For each time step $t$, we use $\mathcal{V}_\mathrm{train}$, $\mathcal{V}_\mathrm{val}$, and $\mathcal{V}_\mathrm{test}$, defined in~\ref{sec:methodology:splits}, to construct year-specific binary node masks. The \textit{training mask} $\mathbf{m}_\mathrm{train}^{(t)}$ is defined as
\begin{equation}
    \mathbf{m}_\mathrm{train}^{(t)}[i] = \begin{cases}
    1 & \text{if } v_i \in \bigcup_{y=t-w}^{t-1}\mathcal{V}_\mathrm{train}^{(y)}, \\
    0 & \mathrm{otherwise,}
\end{cases}
\end{equation}
the \textit{validation mask} $\mathbf{m}_\mathrm{val}^{(t)}$ is defined as
\begin{equation}
    \mathbf{m}_\mathrm{val}^{(t)}[i] = \begin{cases}
    1 & \text{if } v_i \in \mathcal{V}_\mathrm{val}^{(y)}, \\
    0 & \mathrm{otherwise,}
\end{cases}
\end{equation}
and the \textit{test mask} $\mathbf{m}_\mathrm{test}^{(t)}$ is defined as
\begin{equation}
    \mathbf{m}_\mathrm{test}^{(t)}[i] = \begin{cases}
    1 & \text{if } v_i \in \mathcal{V}_\mathrm{test}^{(y)}, \\
    0 & \mathrm{otherwise.}
\end{cases}
\end{equation}

These masks ensure that, for every model, loss and evaluation are computed only for nodes hired in the relevant years, even though all nodes are used for message passing.

\paragraph{Training Setup for Tabular Models}

For each node $v_i$ hired in year $t_i$, we prevent data leakage by using only information prior to the year they are hired as input to the models. We train the neural network models using cross-entropy loss and select the best model checkpoint via the lowest validation loss for evaluation. For the white-box models, we perform grid search with cross-validation. In all cases, the final evaluation is conducted on the test set $\mathcal{V}_\mathrm{test}$. Crucially, all models use the same training and evaluation splits, $\mathcal{V}_\mathrm{train}$, $\mathcal{V}_\mathrm{val}$, and $\mathcal{V}_\mathrm{test}$, as the GNNs, enabling a direct comparison of how different feature types and architectures influence predictive performance.

\paragraph{Training Setup for Static GNNs}

Static graph neural networks like \texttt{GCN}, \texttt{GAT}, and \texttt{GraphSAGE} cannot natively model temporal dynamics. Instead, we treat each test year $t \in \{2018, 2019, 2020\}$ as separate prediction tasks and use a single static graph snapshot $G_{t-1}$ for each task. Node features from year $t-1$, derived from either $\mathbf{X}_\mathrm{PhD}$, $\mathbf{X}_\mathrm{Bib}$, their concatenation, or the constant vector $\mathbf{1}$, are provided as input to the model. For each test year $t$, we reuse the binary node masks defined above to ensure compatibility across models. 

Each model is trained on a snapshot $G_{t-1}$ using cross-entropy loss computed over nodes where $\mathbf{m}_\mathrm{train}^{(t)}[i] = 1$ while validation loss is monitored using $\mathbf{m}_\mathrm{val}^{(t)}$. Model selection is based on early stopping with respect to validation loss. Final performance is computed using predictions on nodes in the test mask $\mathbf{m}_\mathrm{test}^{(t)}$.

All nodes, including those hired prior to $2010$, participate in message passing, but only masked nodes contribute to loss or evaluation. This design ensures that static and temporal models operate under identical node splits and feature inputs, allowing for fair comparison of predictive performance.

\paragraph{Training Setup for Temporal GNNs}

To model the evolving structure of the co-authorship network, \texttt{GConvGRU} is trained using sliding sequences of $w$ consecutive graph snapshots and time-indexed node features. For each year $t$ in ${2010 + w, \dots, 2017}$, the input includes the sequence $\{G_{t-w}, \dots, G_{t-1}\}$ and the corresponding node features (i.e., $\mathbf{X}_\mathrm{PhD}$, $\mathbf{X}_\mathrm{Bib}$, both, or the constant vector $\mathbf{1}$). Masks $\mathbf{m}_\mathrm{train}^{(t)}$, $\mathbf{m}_\mathrm{val}^{(t)}$, and $\mathbf{m}_\mathrm{test}^{(t)}$ restrict loss and evaluation to the relevant nodes for each year, as defined above.

%
% Table 2: Average performance of the best model for each feature set
%
\begin{table*}[t]
\centering
\resizebox{\textwidth}{!}{%
\begin{tabular}{c|c|c|l|c|c|c|c|c}
\multicolumn{1}{l|}{\textbf{PhD}} & \multicolumn{1}{l|}{\textbf{Bib}} & \multicolumn{1}{l|}{\textbf{Co-author}} & \textbf{Optimal Model} & \multicolumn{1}{c|}{\textbf{\begin{tabular}[c]{@{}c@{}}Precision\\(std)\end{tabular}}} & \multicolumn{1}{c|}{\textbf{\begin{tabular}[c]{@{}c@{}}Recall\\(std)\end{tabular}}} & \multicolumn{1}{c|}{\textbf{\begin{tabular}[c]{@{}c@{}}PR-AUC\\(std)\end{tabular}}} & \textbf{\begin{tabular}[c]{@{}c@{}}\% Improvement \\ over PhD Rank\end{tabular}} & \textbf{\begin{tabular}[c]{@{}c@{}}\% Improvement over\\ Avg. Co-author Rank\end{tabular}} \\ \hline \hline
      &                                            &                    &        Random                                  & -                                             & -                                          & $0.222$                                    & $-30.00\%$                                                                              & $-15.59\%$                                                                                        \\
 \checkmark                                        &                                            &                        & PhD Rank                                  & $0.363$ $(0.000)$                             & $0.674$ $(0.000)$                          & $0.317$ $(0.000)$                          & $0.00\%$                                                                                & $20.53\%$                                                                                         \\
       &                                            & \checkmark       & Avg. Co-author Rank                              & $0.392$ $(0.000)$                             & $0.240$ $(0.000)$                          & $0.263$ $(0.000)$                          & $-17.03\%$                                                                              & $0.00\%$                                                                                          \\ \hline
 \checkmark                                        &                                            &                         &   RF                                    & $0.382$ $(0.000)$                             & $0.566$ $(0.000)$                          & $0.345$ $(0.002)$                          & $8.83\%$                                                                                & $31.18\%$                                                                                         \\
      & \checkmark                                 &                           & LR                                     & $0.346$ $(0.000)$                             & $\mathbf{0.705\, (0.000)}$                 & $0.406$ $(0.000)$                          & $28.08\%$                                                                               & $54.37\%$                                                                                         \\
     &                                            & \checkmark                &  GConvGRU                              & $0.394$ $(0.215)$                             & $0.497$ $(0.246)$                          & $0.334$ $(0.032)$                          & $5.36\%$                                                                                & $27.00\%$                                                                                         \\
 \checkmark                                        & \checkmark                                 &                                             &  Transformer           & $\mathbf{0.444\, (0.166)}$                    & $0.193$ $(0.103)$                          & $0.424$ $(0.010)$                          & $33.75\%$                                                                               & $61.22\%$                                                                                         \\
 \checkmark                                        &                                            & \checkmark                &  GConvGRU                              & $0.393$ $(0.051)$                             & $0.694$ $(0.140)$                          & $0.414$ $(0.303)$                          & $30.60\%$                                                                               & $57.41\%$                                                                                         \\
     & \checkmark                                 & \checkmark             &     GraphSAGE                             & $0.350$ $(0.009)$                             & $0.543$ $(0.016)$                          & $0.432$ $(0.012)$                          & $36.28\%$                                                                               & $64.26\%$                                                                                         \\
 \checkmark                                        & \checkmark                                 & \checkmark               &  GAT                                    & $0.369$ $(0.004)$                             & $0.619$ $(0.026)$                          & {\color{blue} $\mathbf{0.458}$ $\mathbf{(0.008)}$}   & {\color{blue} $\mathbf{44.48\%}$}                                                        & {\color{blue} $\mathbf{74.14\%}$}                                                             
\end{tabular}
}
\caption{\textbf{Average performance of the best model for each feature set.} We identify the top-performing model for each feature set based on the average PR-AUC score over $10$ runs. We break down the PR-AUC score into its components, precision and recall, and calculate the percent improvement over the PhD rank and average co-author rank heuristics. Precision and recall are computed using a threshold of $0.5$ on the predicted class probabilities. The proportion of faculty hired at top-$10$ departments is equivalent to the PR-AUC of the random guessing baseline (i.e., $0.222$). The best-performing model by PR-AUC is listed in blue, while the best precision and recall are bolded. There is wide variation in the types of top-performing model across the feature sets, and the overall top performing model is a GAT that utilizes the PhD rank, bibliometric, and co-authorship features.}
\label{tab:best_model_y10}
\end{table*}

%
% Figure 3: Performance across models for PhD rank, bibliometric, and co-authorship feature sets
%
\begin{figure*}[t]
\centering
\includegraphics[width=\textwidth]{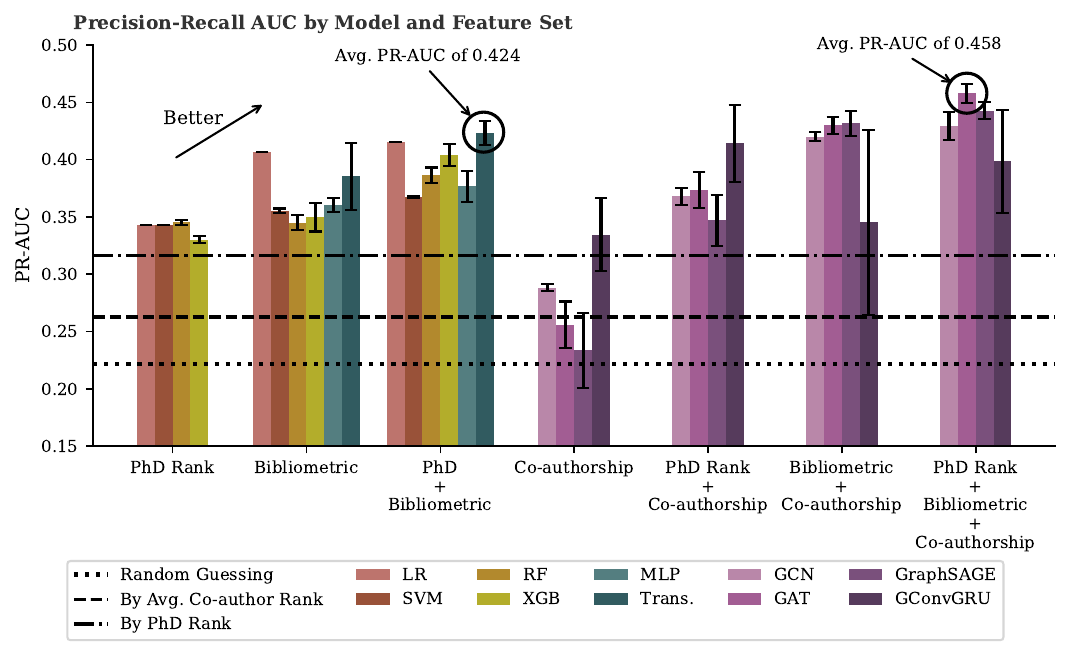}
\caption{\textbf{Predictive performance for top-$\mathbf{10}$ department placement by feature set and model type.} We plot the average PR-AUC and standard deviation across $10$ runs for each of the model trained on our feature sets. Higher values of PR-AUC indicate better performance, and we compare against our three heuristics: random guessing (dotted), the average co-author rank (dashed), and PhD rank (dash-dotted). The performance of the white-box tabular models are depicted in shades of red and yellow, while the neural network-based tabular models are depicted in blues and the graph machine learning models are in shades of purple. We also highlight the top performing graph-based model, a GAT operating on all three feature sets with an average PR-AUC of $0.424$, and the best non-graph model, a transformer which uses PhD rank and bibliometric features to achieve an average PR-AUC of $0.458$. Table~\ref{tab:best_model_y10} contains a full list the best models by feature set. Combining the co-authorship network with meaningful node features leads to the strongest performance.}
\label{fig:bar_chart_y10}
\end{figure*}

At each epoch, the model unrolls through all snapshots, using a recurrent hidden state to propagate information. We track total training and validation loss across all snapshots in the epoch and use the lowest validation loss to select the best model checkpoint for evaluation. During testing, we pass through the full sequence $\{G_{2017}, G_{2018}, G_{2019}\}$, again unrolling the hidden state and performing message passing at each step. The \texttt{GConvGRU} model is optimized using \textit{backpropagation through time}~\cite{werbos.1990} across the sliding window and trained to minimize cross-entropy loss.

\subsection{Modeling Experiment}
\label{sec:SI:experiments}

All models are trained and evaluated using the same temporal train, validation, and test splits described in Sections~\ref{sec:methodology:data} and~\ref{sec:methodology:modeling}. Specifically, the training set includes nodes hired between $2010$ and $2017$, with $80\%$ of each year's hires randomly assigned to the train set and the remaining $20\%$ held out for validation. The test set comprises all nodes hired between $2018$ and $2020$. We perform all experiments under a strict pre-hire constraint: for each node hired in year $t_i$, only data from years prior to year $t_i$ are made available during training or inference.

\paragraph{Tabular Model Hyperparameters}
For tabular models (\texttt{LR}, \texttt{SVM}, \texttt{RF}, \texttt{XGB}, \texttt{MLP}, and \texttt{Trans}), we use Scikit-learn~\cite{scikit} and PyTorch.~\cite{pytorch} For each model, we perform an extensive hyperparameter sweep:
\begin{itemize}
    \item \texttt{LR}: $C \in \{0.0001, 0.001, 0.01, 0.1, 1.0, 10, 100\}$, solvers $\in \{\texttt{liblinear}, \texttt{saga}\}$ and penalties $\in \{\ell_1,\ell_2\}$.
    \item \texttt{SVM}: $C \in \{0.01, 0.1, 1.0, 10, 100\}$, kernels $\in \{\texttt{rbf}, \texttt{linear}\}$, and gamma settings $\in \{\texttt{scale}, \texttt{auto}\}$.
    \item \texttt{RF}: $n_\mathrm{estimators} \in \{100, 200, 500\}$, maximum depth $\in \{5, 10, 20, \mathrm{None}\}$, minimum samples used to split $\in \{2, 5, 10\}$, and maximum number of features $\in \{\texttt{sqrt}, \texttt{log2}, \texttt{None}\}$.
    \item \texttt{XGB}: $n_\mathrm{estimators} \in \{100, 200\}$, maximum depth $\in \{3, 5, 7\}$, learning rate $\in \{0.01, 0.05, 0.1, 0.2\}$, subsample ratios $\in \{0.6, 0.8, 1.0\}$, and minimum samples used to split $\in \{2, 5, 10\}$.
    \item \texttt{MLP} number of layers $\in \{1, 2, 3\}$, hidden channels $\in \{64, 128, 256, 512, 1024, 2048\}$, and activation $\in \{\texttt{ReLU}, \texttt{Tanh}\}$.
    \item \texttt{Trans}: number of layers $\in \{1, 2, 3\}$, hidden channels $\in \{64, 128, 256, 512, 1024, 2048\}$, activation heads $\in \{2, 4, 8\}$, and dropout rates $\in \{0.1, 0.2, 0.3, 0.4, 0.5\}$. 
\end{itemize}
Both the Multi-layer Perceptron and transformer models use a learning rate of $0.001$. We select the best hyperparameter configuration based on PR-AUC and retrain each model $10$ times to estimate variance in test performance.

\paragraph{Graph-model Hyperparameters}

For the graph-based models (\texttt{GCN}, \texttt{GAT}, \texttt{GraphSAGE}, and \texttt{GConvGRU}), we use PyTorch Geometric~\cite{pytorch.geo} and PyTorch Geometric Temporal.~\cite{pytorch.geo.temp} All models use the Adam optimizer~\cite{kingma.2017} with learning rate $0.001$ and are trained for $500$ epochs with early stopping based on validation loss. For each model, we sweep over: number of layers $\in \{1, 2, 3\}$, hidden dimensions $\in \{64, 128, 256, 512, 1024, 2048\}$, dropout rates $\in \{0.1, 0.2, 0.3, 0.4, 0.5\}$, and sliding window widths $w \in \{1, 2, 3\}$. For each graph model and feature set combination, we identify the best hyperparameter configuration using validation PR-AUC, and then retrain the model $10$ times to measure performance robustness under stochastic training conditions.

\section{Results}

We investigate the extent to which faculty placement at high-rank institutions can be predicted from co-authorship networks, by reframing the process as an individual-level, out-of-sample prediction task. Leveraging a dataset of US computer science faculty, we systematically benchmark the predictive value of PhD department prestige, publication-based indicators, and placement in the temporal co-authorship network across a range of model architectures. Our results show that integrating co-authorship structure with traditional features, such as PhD rank or number of first-authored publications, yields the highest predictive accuracy, particularly for distinguishing those who secure positions at the most elite (top-$10$) departments. Notably, the predictive value of co-authorship information declines as the threshold for high-rank is lowered (e.g., from top-$10$ to top-$20$, top-$30$, etc. departments), highlighting the distinctive role of social endorsement and network position in shaping academic career outcomes.

%
% Table 3: Linear mixed-effects model results for predicting placement at top-10 departments
%
\begin{table*}[t]
\centering
\resizebox{\textwidth}{!}{%
\begin{tabular}{lcccccc}
\multicolumn{7}{l}{\textbf{Mixed Linear Model Regression Summary}} \\
\\
\multicolumn{1}{l}{Model:} & \multicolumn{1}{l}{MixedLM} & \multicolumn{1}{l}{Dependent Variable:} & \multicolumn{1}{l}{PR-AUC} & \multicolumn{1}{l}{Method:} & \multicolumn{1}{l}{REML} & \\
\multicolumn{1}{l}{Scale:} & \multicolumn{1}{l}{0.0011} & \multicolumn{1}{l}{Log-Likelihood:} & \multicolumn{1}{l}{912.5417} & \multicolumn{1}{l}{Converged:} & \multicolumn{1}{l}{Yes} & \\
\multicolumn{1}{l}{No. Observations:} & \multicolumn{1}{l}{480} & \multicolumn{1}{l}{No. Groups:} & \multicolumn{1}{l}{10} & \multicolumn{1}{l}{Min. group size:} & \multicolumn{1}{l}{20} & \\
\multicolumn{1}{l}{Max. group size:} & \multicolumn{1}{l}{80} & \multicolumn{1}{l}{Mean group size:} & \multicolumn{1}{l}{48.0} &  &  & \\
& & & & & & \\
\multicolumn{7}{l}{\textbf{Reference = PhD Rank}} \\
Term & Coef. & Std. Err. & $z$ & $P>|z|$ & CI Lower & CI Upper \\
\hline
Intercept & $0.342$ & $0.007$ & $45.797$ & $0.000$ & $0.327$ & $0.357$ \\
PhD Rank $+$ Co-authorship & $0.029$ & $0.010$ & $2.828$ & $0.005$ & $0.009$ & $0.049$ \\
Group Variance & $0.000$ & $0.003$ &  &  & &  \\
& & & & & & \\
& & & & & & \\
\multicolumn{7}{l}{\textbf{Reference = Bibliometric}} \\
Term & Coef. & Std. Err. & $z$ & $P>|z|$ & CI Lower & CI Upper \\
\hline
Intercept & $0.367$ & $0.007$ & $56.021$ & $0.000$ & $0.354$ & $0.380$ \\
Bibliometric $+$ Co-authorship & $0.037$ & $0.010$ & $3.827$ & $0.000$ & $0.018$ & $0.056$ \\
Group Variance & $0.000$ & $0.003$ &  &  & &  \\
& & & & & & \\
\multicolumn{7}{l}{\textbf{Reference = PhD Rank $+$ Bibliometric}} \\
Term & Coef. & Std. Err. & $z$ & $P>|z|$ & CI Lower & CI Upper \\
\hline
Intercept & $0.396$ & $0.007$ & $60.374$ & $0.000$ & $0.383$ & $0.408$ \\
PhD Rank $+$ Bibliometric $+$ Co-authorship & $0.029$ & $0.010$ & $3.005$ & $0.003$ & $0.010$ & $0.048$ \\
Group Variance & $0.000$ & $0.003$ &  &  & &  \\
\end{tabular}
}
\caption{\textbf{Linear mixed-effects model results for predicting placement at top-$\mathbf{10}$ departments.} Summary of mixed-effects regression models predicting PR-AUC as a function of feature set (fixed effect) and model architecture (random effect). The figure reports model diagnostics, including the number of observations and groups, group size range and mean, and estimation method. We consider three reference feature sets (intercepts): (1) PhD rank, (2) bibliometric features, and (3) both PhD rank and bibliometric features. For each, we report the coefficient, standard error, $z$-score, $p$-value, and $95\%$ confidence interval associated with adding co-authorship features. Regression coefficients for additional feature sets are provided in Table~\ref{tab:stats_y10}. In all cases, incorporating co-authorship information yields a statistically significant improvement in PR-AUC, despite variation across model architectures.}
\label{tab:stats_y10_rewired}
\end{table*}

\subsection{Incorporating co-authorship structure significantly improves faculty placement prediction}
\label{sec:results:incorporating-coauthorship}

Predicting where early-career researchers will be hired is a challenging task, especially given that most candidates have relatively short publication track records and frequently share similar markers of institutional prestige. We evaluate a comprehensive set of machine learning models (see Sections~\ref{sec:results:task_definition},~\ref{sec:methodology:modeling}, and Tab.~\ref{tab:models}) trained on various combinations of pre-hire features, including PhD department rank, bibliometric indicators (such as publication and citation counts), and co-authorship network structure, to assess which factors provide meaningful predictive signal of whether a faculty candidate is hired at a high-rank department. Here, we define \textit{high-rank} to mean \textit{a top-$10$ department}. Different definitions of \textit{high-rank} are explored in Section~\ref{sec:results:robustness}. We benchmark performance against three simple heuristics: random guessing, and predicting researchers are hired at high-rank departments if their PhD department is high-rank or if the average prestige of their co-authors's departments is high-rank.

Our main finding is that integrating co-authorship network structure with meaningful node features significantly improves predictive performance, outperforming both established heuristics and models using traditional features alone. As summarized in Table~\ref{tab:best_model_y10}, the top-performing model, a Graph Attention Network (GAT) trained on the full feature set (PhD rank, bibliometric features, and the co-authorship network), achieves a PR-AUC of $0.458$, representing a $44.48\%$ improvement over the PhD rank heuristic and a $74.14\%$ gain over the average co-author rank heuristic. Importantly, the three feature sets that combine co-authorship with either PhD rank, bibliometric data, or combined PhD rank and bibliometric features rank within the top four by predictive performance. The remaining spot is occupied by the combined PhD rank and bibliometric feature set (without co-authorship features). This pattern underscores the consistent contribution of co-authorship information to the most effective models.

Examining the baseline and single-feature models clarifies the limitations of traditional predictors. Models trained only on PhD department rank or bibliometric data outperform baselines for their best-performing configurations (Random Forest achieves a PR-AUC of $0.345$ with PhD rank, while Logistic Regression achieves $0.406$ with bibliometric features), but do not consistently surpass the heuristics across all models and runs (see Fig.~\ref{fig:bar_chart_y10}). When PhD rank and bibliometric features are combined, average model performance improves with a transformer leading to the best average PR-AUC of $0.424$, but substantial variability remains and these models still trail those that incorporate co-authorship information (see Table~\ref{tab:best_model_y10}). The bar chart in Figure~\ref{fig:bar_chart_y10} highlights this trend, illustrating that feature sets containing co-authorship information in addition to the other features consistently yield higher and more consistent PR-AUC scores across a range of model architectures compared to the tabular feature-only models.

These patterns reflect two key realities of academic hiring. First, candidates' short career ages limit the differentiation possible from productivity-based metrics alone. Second, a well-established prestige hierarchy, like the one discussed by Clauset et al.~\cite{clauset.2015}, results in many candidates with similarly strong institutional backgrounds competing for a limited number of high-rank positions. As a result, even features that are strongly correlated with overall prestige at a population level offer limited power to distinguish among top candidates.

%
% Table 4
%
\begin{table*}[t]
\centering
\resizebox{\textwidth}{!}{%
\begin{tabular}{c|l|c|c|l}
\multicolumn{1}{l|}{\textbf{Threshold}} & \textbf{Optimal model} & \multicolumn{1}{l|}{\textbf{PR-AUC (std)}} & \multicolumn{1}{l|}{\textbf{Prop. High-rank Hires}} & \textbf{Features}     \\ \hline \hline
Top-$10$                                & GAT                    & $0.458$ $(0.008)$                          & $0.222$                                             & PhD + Bib + Co-author \\
Top-$20$                                & GraphSage              & $0.589$ $(0.009)$                          & $0.357$                                             & PhD + Bib + Co-author \\
Top-$30$                                & GConvGRU               & $0.712$ $(0.039)$                          & $0.491$                                             & PhD + Co-author       \\
Top-$40$                                & GCN                    & $0.846$ $(0.003)$                          & $0.674$                                             & PhD + Bib + Co-author \\
Top-$50$                                & GCN                    & $0.914$ $(0.001)$                          & $0.771$                                             & PhD + Bib + Co-author
\end{tabular}
}
\caption{\textbf{The top-performing models for different thresholds of \textit{high-rank}.} For each definition of high-rank, we identify the best-performing model and feature set based on the average PR-AUC score over $10$ runs. Because PR-AUC depends on the proportion of positive class labels, the random baseline and, consequently, the model performance increases with the proportion of high-rank hires. In all cases, the top-performing models are graph-based models and, with the exception of top-$30$, the models use a combination of PhD rank, bibliometric, and co-authorship features.}
\label{tab:best_models}
\end{table*}
%
% Figure4: The impact of the co-authorship network on performance and the usefulness of PhD rank vs. bibliometric features for different thresholds for high-rank
%
\begin{figure*}[t]
\centering
\includegraphics[width=\textwidth]{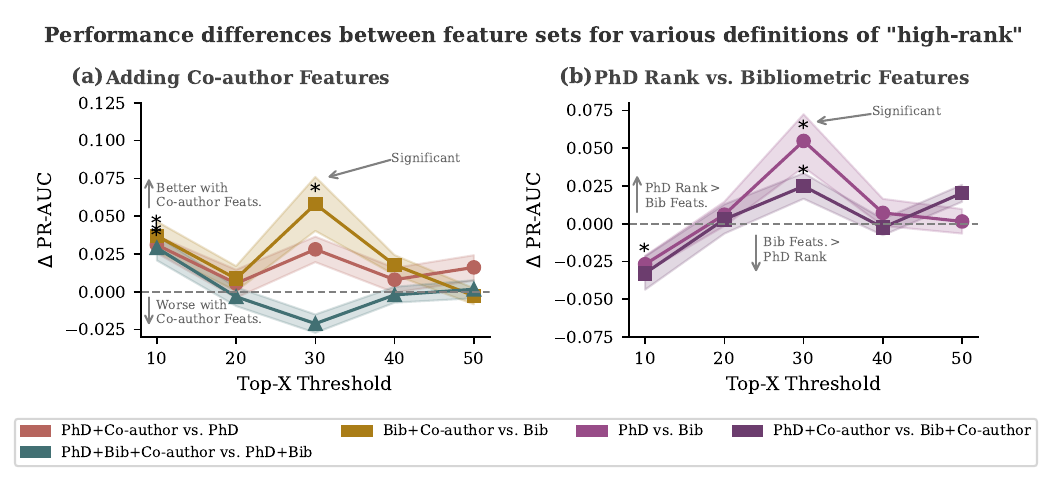}
\caption{\textbf{The impact of the co-authorship network on performance and the usefulness of PhD rank vs. bibliometric features for different thresholds for \textit{high-rank}.} Each line shows the average difference in PR-AUC (with $90\%$ confidence intervals) between two feature sets. In \textbf{(a)}, comparisons involve adding co-authorship features to either PhD rank (red line), bibliometric features (yellow line), or both (blue line); Lines above the dashed line ($\Delta$PR-AUC$ > 0$) indicate an improvement. In \textbf{(b)}, comparisons evaluate feature sets with PhD rank vs. bibliometric features with (purple line) and without (pink line) co-authorship features; Lines above the dashed line favor PhD rank. Asterisks mark statistically significant differences ($p < 0.05$; see Tables~\ref{tab:stats_y10},~\ref{tab:stats_y20}-\ref{tab:stats_y50}). As the definition of \textit{high-rank} broadens (i.e., from top-$10$ to top-$50$), the benefit of co-authorship features declines, while PhD rank becomes more predictive.}
\label{fig:robustness}
\end{figure*}

Incorporating co-authorship network structure introduces a new, socially meaningful axis of differentiation. While co-authorship information alone does not yield strong predictive signals, its value emerges when combined with node-level features. Notably, these network ties may function as both explicit and implicit signals of social endorsement—explicitly, when co-authorship reflects collaboration or mentorship with prominent researchers (serving as a proxy for recommendation writers), and implicitly, by signaling broader integration or recognition within the research community. By fusing these social signals with traditional metrics, our models are better able to identify promising candidates when other indicators are ambiguous.

The predictive gains from incorporating co-authorship structure are robust and statistically significant. Figure~\ref{fig:bar_chart_y10} shows the model-level PR-AUC scores across feature sets, revealing that models using both the co-authorship network and node features achieve the highest median and most consistent performance. The combined distributions of PR-AUC scores by feature set can be found in Figure~\ref{fig:box_y10}. Linear mixed-effects regression further confirms these effects: including the co-authorship network leads to a $8.48\%$ ($p=0.005$) increase in PR-AUC over PhD rank alone, a $10.08\%$ ($p<0.001$) increase over bibliometric features alone, and a $7.32\%$ ($p=0.003$) increase over both combined (Table~\ref{tab:stats_y10_rewired}).

Despite these improvements, even the best models achieve PR-AUC below $0.5$, underscoring the inherent noise and complexity of faculty hiring decisions. Many relevant factors, including letters of recommendation, subfield-specific considerations, and idiosyncratic aspects of evaluation, remain unobserved. Nevertheless, our results demonstrate that combining co-authorship structure with node-level features not only improves predictive accuracy, but also provides insight into the subtle mechanisms by which social context and institutional signals jointly shape academic career outcomes.

\subsection{Co-authorship structure is most informative for predicting top-tier placements}
\label{sec:results:robustness}

While the previous section demonstrated that productivity-based and PhD rank prestige features alone are often insufficient for distinguishing the very best faculty placements, those findings relied on defining \textit{high-rank} as placement at a top-$10$ department. However, only about $22\%$ of early-career faculty are hired at top-$10$ departments, while $36\%$ are hired at top-$20$, $49\%$ at top-$30$, $67\%$ at top-$40$, and $77\%$ are hired at top-$50$ departments. Here, we test whether our definition of high-rank shapes our conclusions by systematically repeating our analysis with different thresholds: top-$20$, top-$30$, top-$40$, and top-$50$ departments. This allows us to assess whether the value of co-authorship information and other features depends on how narrowly we define elite appointments. Figures~\ref{fig:bar_chart_y20_rewiring}-\ref{fig:box_y50} and Tables~\ref{tab:best_model_y20}-\ref{tab:best_model_y50} contain full performance breakdowns at each threshold.

Table~\ref{tab:best_models} summarizes the top-performing model and feature set for each threshold. Across all thresholds, the best-performing models are graph-based, indicating that relational information provides consistent value regardless of how elite placement is defined. In all but one case, the best feature set includes the full combination of PhD rank, bibliometric, and co-authorship features. The exception is the top-$30$ prediction task, where the optimal model (GConvGRU) uses only PhD rank and co-authorship structure. This shift in feature importance may reflect the approximately balanced class distribution for the top-$30$ threshold ($49.1\%$ high-rank), which could help graph neural networks perform better without requiring the additional bibliometric features. Alternatively, it may indicate that bibliometric features provide less marginal value for distinguishing candidates at this level of selectivity. The optimal models for the remaining feature sets are listed in Tables~\ref{tab:best_model_y10} and~\ref{tab:best_model_y20}-\ref{tab:best_model_y50}.

If we look beyond the top-performing model, we see that the predictive advantage of co-authorship structure diminishes as the definition of \textit{high-rank} broadens. When high-rank is defined narrowly (e.g., top-$10$), adding co-authorship features yields statistically significant improvements across all models (see Fig.~\ref{fig:robustness}(a); statistical details in Tables~\ref{tab:stats_y10},~\ref{tab:stats_y20}-\ref{tab:stats_y50}). However, these gains taper off at broader thresholds, where co-authorship features provide less additional signal. While the best overall models are graph-based models, this pattern suggests that, on average, co-authorship networks are most informative when distinguishing among candidates who are otherwise similarly elite, i.e., those competing for the most selective positions.

In contrast, PhD rank becomes an increasingly strong predictor relative to bibliometric features as the definition of high-rank broadens. PhD rank-based feature sets outperform bibliometric-based feature sets starting at the top-$20$ threshold (see Fig.\ref{fig:robustness}(b)). This pattern continues through the top-$50$ threshold, reinforcing the idea that doctoral pedigree becomes more useful for identifying those \textit{unlikely} to reach the top, even if it does not separate those already at the top of the pool. Notably, both Figures~\ref{fig:robustness}(a) and~\ref{fig:robustness}(b) have outliers in the difference in performance of the feature sets for the top-$30$ threshold, again pointing to the possible influence of data balance at that cutoff.

Taken together, these findings reveal a clear pattern: on average, co-authorship structure is most useful for differentiating among the most elite candidates---those who might otherwise appear indistinguishable in terms of productivity or prestige---while PhD rank becomes increasingly predictive as we expand the definition of high-rank to include a broader pool. Further, graph-based models consistently achieve the best performance, underscoring the importance of social structure in predicting academic mobility, especially at the upper echelons of the prestige hierarchy.

\section{Discussion}

This study reframes the prediction of faculty placement as an individual-level, out-of-sample prediction problem and finds that a candidate’s position in the temporal co-authorship network, when combined with traditional features such as PhD rank and bibliometric attributes, significantly enhances predictive accuracy. Our analysis across nearly $2000$ US computer science faculty demonstrates that the social structure encoded by co-authorship patterns provides a measurable and unique axis of differentiation among highly credentialed early-career researchers. This effect is especially pronounced at the very top of the academic hierarchy, where traditional markers such as PhD department rank and publication record offer limited ability to distinguish among otherwise similar candidates.

These findings carry several important implications. First, they offer quantitative support for the long-held intuition that scholarly success and mobility are shaped not only by individual accomplishment but also by a researcher’s integration within broader intellectual and social communities. The fact that co-authorship features are most predictive for the most elite placements suggests that social endorsement, advocacy, and network-based reputation may play a particularly strong role when committees are faced with a pool of highly qualified candidates. Graduating from a highly ranked PhD program appears to be a necessary---but not a sufficient---condition for securing a top faculty appointment. The predictive value of PhD rank is limited among the most elite outcomes due to the sheer abundance of highly qualified applicants. As the definition of high-rank expands to include a larger set of departments, PhD rank becomes increasingly effective at identifying those who are \textit{less likely} to secure top placements. This pattern aligns with longstanding findings on the role of prestige hierarchies in academic hiring~\cite{clauset.2015, burris.2004}, but also underscores a crucial distinction: while prestige hierarchies are powerful for explaining group-level trends, they are less useful for distinguishing among those who already have access to the advantages conferred by prestigious academic affiliations.

Second, our work highlights both the potential and the challenges of using network-derived features to understand academic prestige hierarchies. While our results show that co-authorship structure encodes meaningful information visible to hiring committees, the overall predictive performance remains modest (with PR-AUC scores below $0.5$ even for the best models when predicting top-$10$ hires). This underscores the inherent complexity and noise in faculty hiring outcomes, and points to a number of unmeasured factors, such as recommendation letter writers, subfield-specific hiring dynamics, and informal mentoring relationships, that remain outside the scope of most available datasets.

\paragraph{Limitations}

Our study has several limitations. First, it focuses on US computer science faculty and relies on DBLP publication records and reputation-based rankings, which may limit generalizability to other disciplines, countries, or non-tenure-track appointments. Second, while co-authorship is a useful proxy for professional relationships, it cannot capture all forms of social capital or informal advocacy, and may conflate different types of collaboration. Third, our models condition on eventual hiring and do not address who pursues or is offered faculty positions. We also discretize placement prestige into bins rather than modeling it as a continuous outcome. While this decision is informed by previous work on prestige hierarchies, it may obscure finer-grained patterns. Finally, because we fit separate models for each feature set, some performance differences may arise from architectural or hyperparameter differences rather than the features alone. While we partially address this using linear mixed-effects modeling, some structural differences in the data remain. Lastly, our machine learning approach is predictive, not causal: we do not attempt to disentangle the effects of correlated features, such as the influence of PhD department on one’s co-author network. Thus, observed associations should not be interpreted as evidence of direct causal relationships.

\paragraph{Future Work}
Looking ahead, future research can address these limitations in several ways. Expanding this predictive framework to other fields or global datasets may reveal both generalities and domain-specific differences in the role of co-authorship networks in faculty placement. Enhancing the underlying data, such as by incorporating measures of publication quality (e.g., citation counts or journal prestige), could provide a more nuanced account of scholarly achievement and help disentangle the influence of network position from that of research impact. Integrating additional sources of social context, including mentorship ties, conference participation, or explicit recommendation letter networks, could further improve model accuracy and illuminate the often-invisible scaffolding that supports academic mobility.

Further, pairing these network-based analyses with demographic attributes would enable a deeper examination of whether the observed biases and advantages associated with co-authorship networks translate into systematic inequities. If so, this line of work points toward a promising avenue for intervention: unlike many systemic reforms, co-authorship-based interventions can operate at the individual level, potentially empowering researchers to expand their opportunities through intentional collaboration. Such strategies may offer more accessible approaches to mitigating bias than structural or policy-based measures alone. More broadly, examining the interplay between structural advantage, network evolution, and institutional decision-making could help untangle the mechanisms that perpetuate or disrupt academic hierarchies. Ultimately, these methods can inform efforts to democratize access to academic opportunity by making visible the often-hidden channels through which social capital and endorsement shape career trajectories.

\section{Conclusion}
Although faculty hiring is often framed as a meritocratic assessment of individual accomplishment, our findings show that a candidate’s social positioning, captured through temporal co-authorship networks, can meaningfully predict the prestige of their first faculty appointment. By reframing faculty placement as an individual-level prediction task and systematically evaluating multiple forms of pre-hire information, we demonstrate that network structure offers complementary insight beyond what is captured by PhD rank and productivity measures alone. This predictive signal is especially valuable at the uppermost tiers of the academic hierarchy, where traditional credentials fail to differentiate among top candidates.

More broadly, our work highlights how social capital and informal endorsement, factors often hidden from formal metrics, can shape early-career outcomes in measurable ways. In particular, co-authorship networks may serve as proxies for informal advocacy or recommendation letter writers, offering hiring committees a glimpse into a candidate’s embeddedness in influential academic circles. By making these dynamics visible, this study provides new tools for interrogating fairness in academic hiring and raises the possibility of designing interventions that operate at the institutional level through intentional collaboration and network-building. Understanding how professional proximity influences opportunity is critical for promoting transparency and equity in access to top academic positions.

\section*{Acknowledgments}

We thank Sanjukta Krishnagopal, Maximilian Nickel, Sam Zhang, Germans Savcisens, Zohair Shafi, and Moritz Laber for their feedback and stimulating discussions on methodological and empirical portions of this work.

\section*{Data Availability}

All data used in this analysis are publicly available. The raw data for the DBLP co-authorship is available at~\url{https://dblp.org/}, the CSRankings data is available at~\url{https://csrankings.org/}, and the CS Professors data is available at~\url{https://jeffhuang.com/computer-science-open-data/}.

\section*{Code Availability}

The code used to preprocess the data, train the models, evaluate the models, and generate the paper figures is publicly available at~\url{https://github.com/samanthadies/predicting-faculty-placement}. The authors used generative AI to help style portions of the figures, prepare the data, and generate first drafts of pieces of the modeling scripts. All code generated or modified by AI was checked for correctness.

\bibliography{refs.bib}

\clearpage

\FloatBarrier
\begin{appendices}
\onecolumn

\setcounter{figure}{0}
\setcounter{table}{0}
\renewcommand{\thefigure}{A\arabic{figure}}
\renewcommand{\thetable}{A\arabic{table}}

\section{Rewiring Co-authorship Networks}
\label{sec:SI:rewiring}

To evaluate the degree to which the specific edges in the co-authorship network carry meaningful predictive signals beyond degree, we train our models on rewired versions of the co-authorship network. Specifically, we iteratively rewire edges in our co-authorship network to generate versions which are $10\%$ rewired, $20\%$ rewired, and so on until we reach a fully random network in which the degree of each node is preserved but all other structural information is randomized. This procedure allows us to evaluate whether higher-order structural signals, such as triadic closure, clustering, or brokerage, contribute to prediction beyond what can be inferred from degree alone. 

The degree-preserving randomization using the double-edge swap algorithm, which repeatedly selects two edges and rewires them while preserving node degrees. In a static network, this involves choosing two random edges, $(v_1, v_2)$ and $(v_3, v_4)$, and rewiring them to $(v_1, v_3)$ and $(v_2, v_4)$, provided that neither of the new edges already exists. In our temporal co-authorship, however, node degrees change across the snapshots. Therefore, in order to preserve degree in each year $t$, we rewire each snapshot by rewiring only the edges that are introduced in each subsequent snapshot. First, we store all edge sets $\mathcal{E} = \{\mathcal{E}_{2010}, \dots, \mathcal{E}_{2020}\}$. These edge sets are cumulative. Then, for each year $t \in \{2010, \dots, 2020\}$, we rewire $p\%$ of the edges $\mathcal{E}_\mathrm{non\_cumulative}^{(t)} = \mathcal{E}_t \setminus \mathcal{E}_{t-1}$ for $p \in \{10, 20, ..., 90, 100\}$. After the non-cumulative edge sets are rewired, we re-aggregate them into new, randomized graph sequences $\tilde{G}^{p}=\{\tilde{G}^{p}_{2010}, \dots, \tilde{G}^{p}_{2020}\}$.

During training, we train \texttt{GConvGRU}, \texttt{GCN}, \texttt{GAT}, and \texttt{GraphSAGE} with the same node masks, loss functions, and hyperparameters discussed in Sections~\ref{sec:methodology} and~\ref{sec:SI:experiments}. The only two differences are the input graphs, where we substitute $\tilde{G}^{p}_t$ for $G_t$. This setup allows us to quantify the performance gain from higher-order structural signals in the co-authorship network. Rather than tuning new hyperparameters, each graph model is trained on $\tilde{G}^{p}$ using the best configuration found in the original (non-rewired) experiments. We generate $10$ independent rewired graph sequences for each rewiring percentage $p$ and train one model per sequence, yielding $10$ runs per model, rewiring percentage, and feature set. This mirrors the stochasticity estimation procedure used in the tabular and graph-based models, while preserving the controlled randomness introduced through rewiring.

\begin{figure}[t]
\centering
\includegraphics[width=\textwidth]{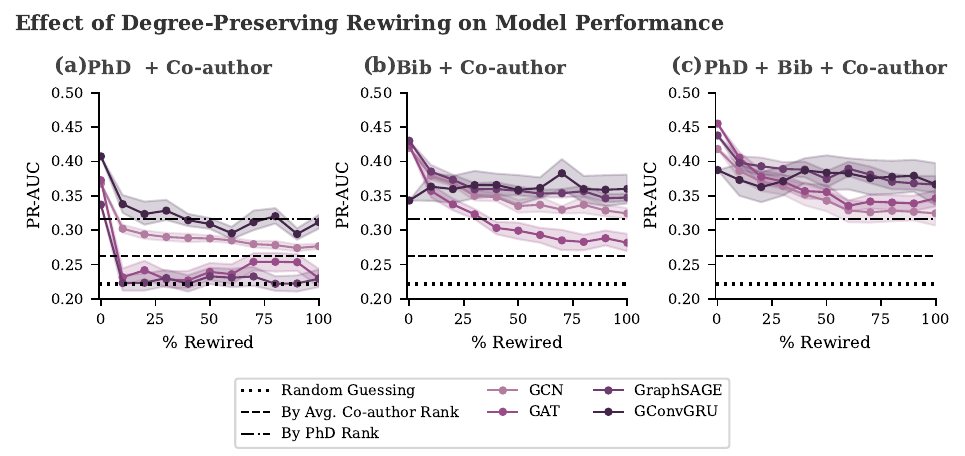}
\caption{\textbf{Predictive performance for top-$\mathbf{10}$ department placement of graph-based models as co-authorship edges are progressively rewired.} We plot the PR-AUC (with 90\% confidence intervals) for the graph-based models, GCN, GAT, GraphSAGE, and GConvGRU, as the edges in the co-authorship network are increasingly randomized using degree-preserving randomization. Models are trained using three feature set combinations: \textbf{(a)} PhD rank and co-authorship, \textbf{(b)} bibliometric and co-authorship, and \textbf{(c)} all three combined. Higher PR-AUC indicates better predictive performance. Horizontal lines show theheuristic baselines: random guessing (dotted), average co-author rank (dashed), and PhD rank (dash-dotted).}
\label{fig:rewiring}
\end{figure}

As demonstrated in Figure~\ref{fig:rewiring}, the performance of our graph-based models drops as the percent of rewired edges increases. In the case of the PhD rank and co-authorship network feature combination, performance drops heavily and levels off at a PR-AUC of around $0.33$ for GConvGRU, $0.30$ for GCN, $0.25$ for GAT, and $0.22$ for GraphSage (see Fig.~\ref{fig:rewiring}(a)). We also see sharp declines in performance between the original network and $\tilde{G}^{10}$ for the bibliometric and co-authorship feature combination (Fig.~\ref{fig:rewiring}(b)) and the full combined feature set (Fig.~\ref{fig:rewiring}(c)), with continued but more gradual declines as the percentages of rewired edges increase. These results demonstrate that higher-order structure in co-authorship networks contributes to prediction beyond what is captured by node features or degree alone. That even partially rewired graphs degrade performance highlights the importance of preserving relational context in modeling faculty placement.

%
% rewiring plot
%

\section{Supplementary Top-\texorpdfstring{$\mathbf{10}$}{10} Results}
\label{sec:SI:top_10_results}

\begin{figure}[t]
\centering
\includegraphics[width=.9\textwidth]{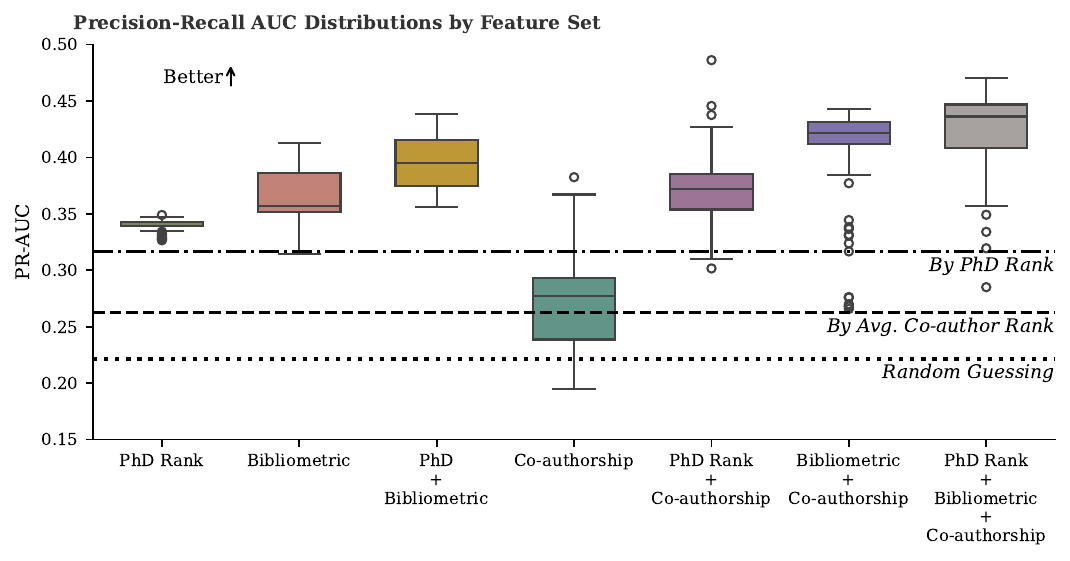}
\caption{\textbf{Predictive performance for top-$\mathbf{10}$ department placement grouped by feature set.} We plot comparative boxplots of the distribution of PR-AUC scores for the models trained on each combination of our feature sets. The performance breakdown by model can be found in Figure~\ref{fig:bar_chart_y10}. Higher values of PR-AUC indicate better performance, and we compare against our three heuristics: random guessing (dotted), the average co-author rank (dashed), and PhD rank (dash-dotted).}
\label{fig:box_y10}
\end{figure}

Here we present additional results on model performance for predicting whether prospective faculty are hired at top-$10$ departments. As discussed in Section~\ref{sec:results:incorporating-coauthorship}, combining co-authorship features with PhD rank and bibliometric indicators yields the strongest performance overall (see Fig.~\ref{fig:bar_chart_y10}, Tab.~\ref{tab:best_model_y10}). Figure~\ref{fig:box_y10} shows an alternative view of these results by aggregating performance across all models to generate box plots at the feature set level. These plots illustrate the distribution of PR-AUC scores for each feature set combination. As expected, models using all three types of feature sets, PhD rank, bibliometric, and co-authorship, achieve the highest performance on average. Finally, Table~\ref{tab:stats_y10} reports the full statistical tests supporting these findings. In all comparisons, adding co-authorship features leads to a statistically significant improvement in PR-AUC.

\section{Supplementary Results for Broadened Definitions of \textit{High-Rank}}
\label{sec:SI:full_results}

\begin{figure}[t]
\centering
\includegraphics[width=\textwidth]{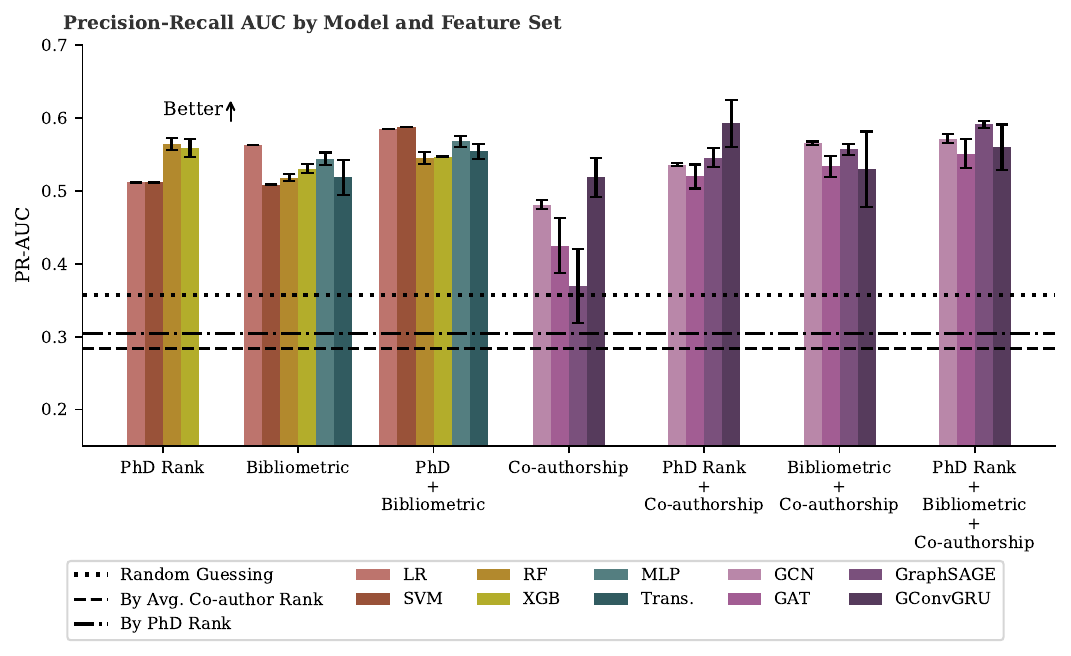}
\caption{\textbf{Predictive performance for top-$\mathbf{20}$ department placement by feature set and model type.} We plot the average PR-AUC and standard deviation across $10$ runs for each of the model trained on our feature sets. Higher values of PR-AUC indicate better performance, and we compare against our three heuristics: random guessing (dotted), the average co-author rank (dashed), and PhD rank (dash-dotted).}
\label{fig:bar_chart_y20_rewiring}
\end{figure}

\begin{figure}[t]
\centering
\includegraphics[width=.94
\textwidth]{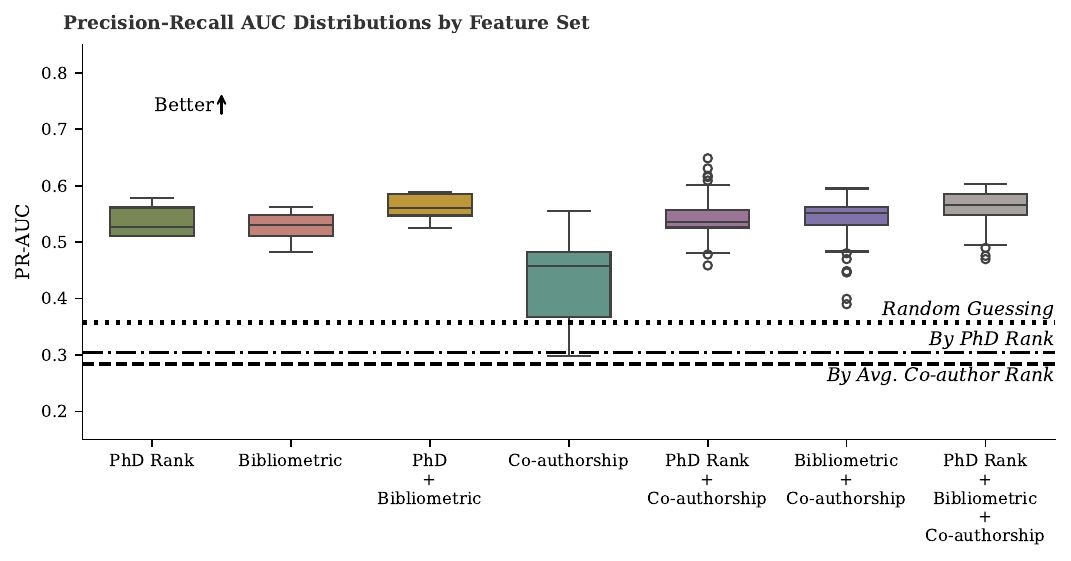}
\caption{\textbf{Predictive performance for top-$\mathbf{20}$ department placement grouped by feature set.} We plot comparative boxplots of the distribution of PR-AUC scores for the models trained on each combination of our feature sets. The performance breakdown by model can be found in Figure~\ref{fig:bar_chart_y20_rewiring}. Higher values of PR-AUC indicate better performance, and we compare against our three heuristics: random guessing (dotted), the average co-author rank (dashed), and PhD rank (dash-dotted).}
\label{fig:box_y20}
\end{figure}

Here we report the full set of results for model performance as the definition of high-rank is expanded to include top-$20$, $30$, $40$, and $50$ departments. For the top-$20$ case, model-level performance across feature sets is shown in Figure~\ref{fig:bar_chart_y20_rewiring}, while Table~\ref{tab:best_model_y20} lists the best-performing model for each feature set. Figure~\ref{fig:box_y20} shows the distribution of performance across models. As with the top-$10$ results, the best-performing model combines PhD rank, bibliometric, and co-authorship features. However, differences in performance between feature sets are notably smaller, both in the best-case and average-case.

\begin{figure}[t]
\centering
\includegraphics[width=.9\textwidth]{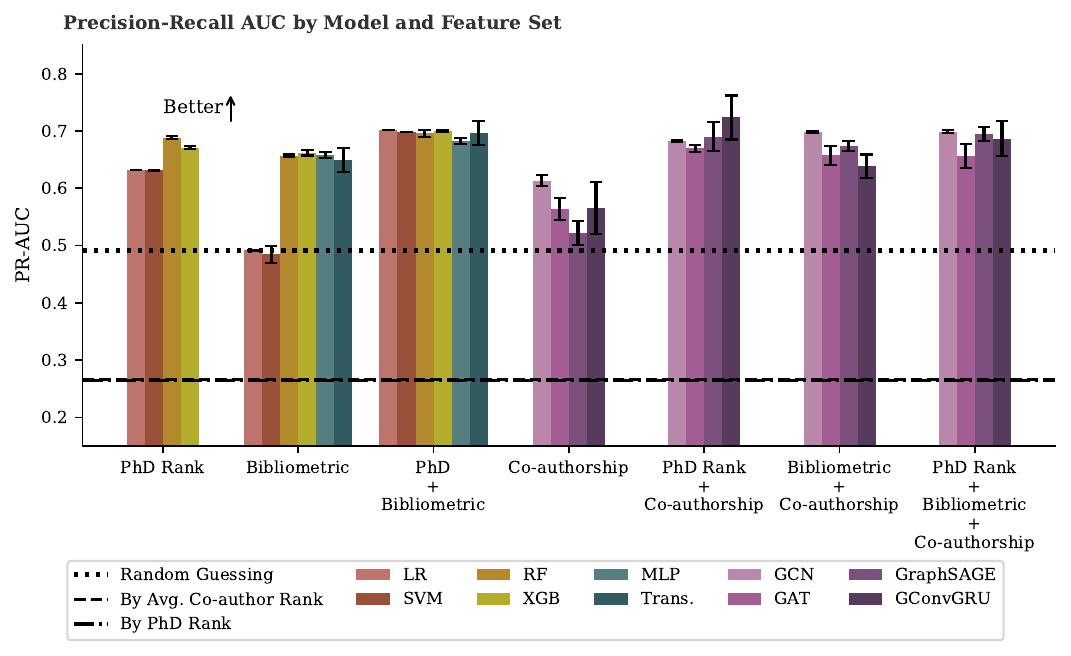}
\caption{\textbf{Predictive performance for top-$\mathbf{30}$ department placement by feature set and model type.} We plot the average PR-AUC and standard deviation across $10$ runs for each of the model trained on our feature sets. Higher values of PR-AUC indicate better performance, and we compare against our three heuristics: random guessing (dotted), the average co-author rank (dashed), and PhD rank (dash-dotted).}
\label{fig:bar_chart_y30_rewiring}
\end{figure}

\begin{figure}[t]
\centering
\includegraphics[width=.9\textwidth]{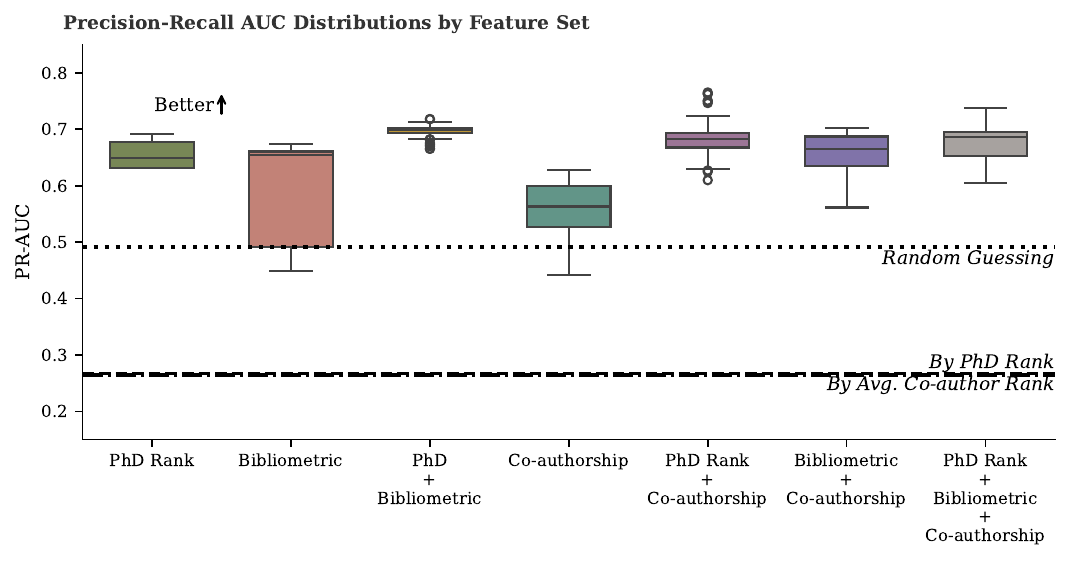}
\caption{\textbf{Predictive performance for top-$\mathbf{30}$ department placement grouped by feature set.} We plot comparative boxplots of the distribution of PR-AUC scores for the models trained on each combination of our feature sets. The performance breakdown by model can be found in Figure~\ref{fig:bar_chart_y30_rewiring}. Higher values of PR-AUC indicate better performance, and we compare against our three heuristics: random guessing (dotted), the average co-author rank (dashed), and PhD rank (dash-dotted).}
\label{fig:box_y30}
\end{figure}

\begin{figure}[t]
\centering
\includegraphics[width=.9\textwidth]{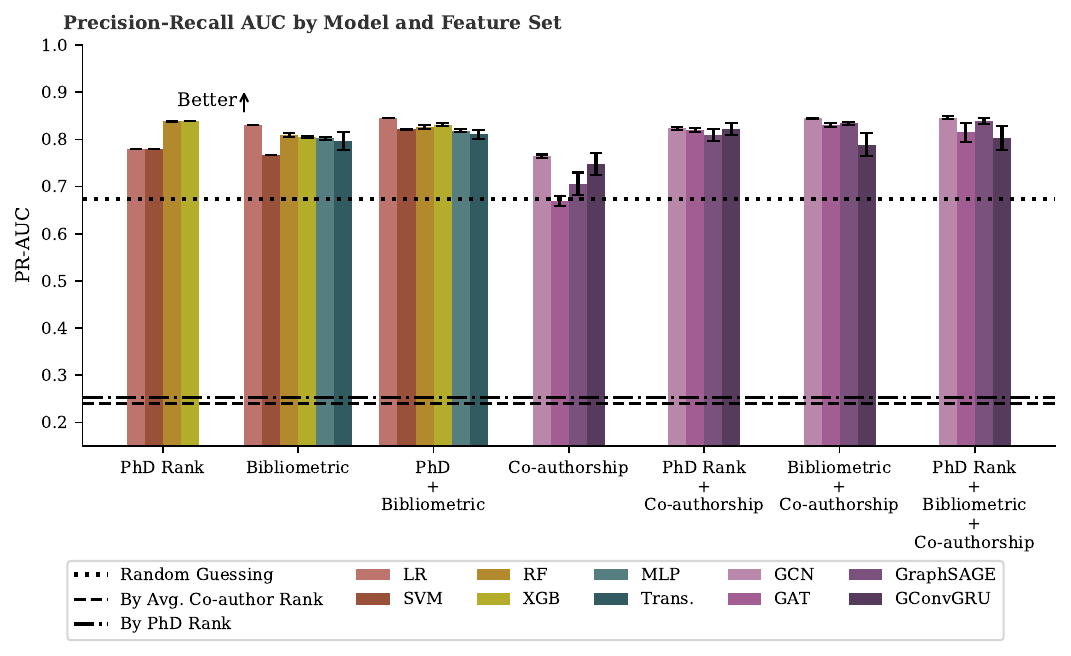}
\caption{\textbf{Predictive performance for top-$\mathbf{40}$ department placement by feature set and model type.} We plot the average PR-AUC and standard deviation across $10$ runs for each of the model trained on our feature sets. Higher values of PR-AUC indicate better performance, and we compare against our three heuristics: random guessing (dotted), the average co-author rank (dashed), and PhD rank (dash-dotted).}
\label{fig:bar_chart_y40_rewiring}
\end{figure}

\begin{figure}[t]
\centering
\includegraphics[width=.9\textwidth]{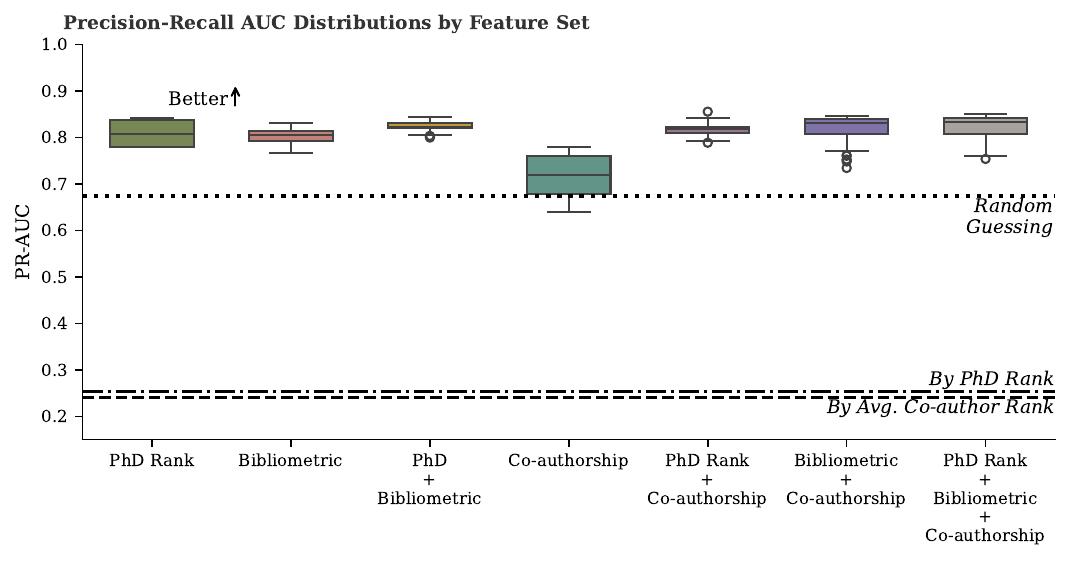}
\caption{\textbf{Predictive performance for top-$\mathbf{40}$ department placement grouped by feature set.} We plot comparative boxplots of the distribution of PR-AUC scores for the models trained on each combination of our feature sets. The performance breakdown by model can be found in Figure~\ref{fig:bar_chart_y40_rewiring}. Higher values of PR-AUC indicate better performance, and we compare against our three heuristics: random guessing (dotted), the average co-author rank (dashed), and PhD rank (dash-dotted).}
\label{fig:box_y40}
\end{figure}

\begin{figure}[t]
\centering
\includegraphics[width=.9\textwidth]{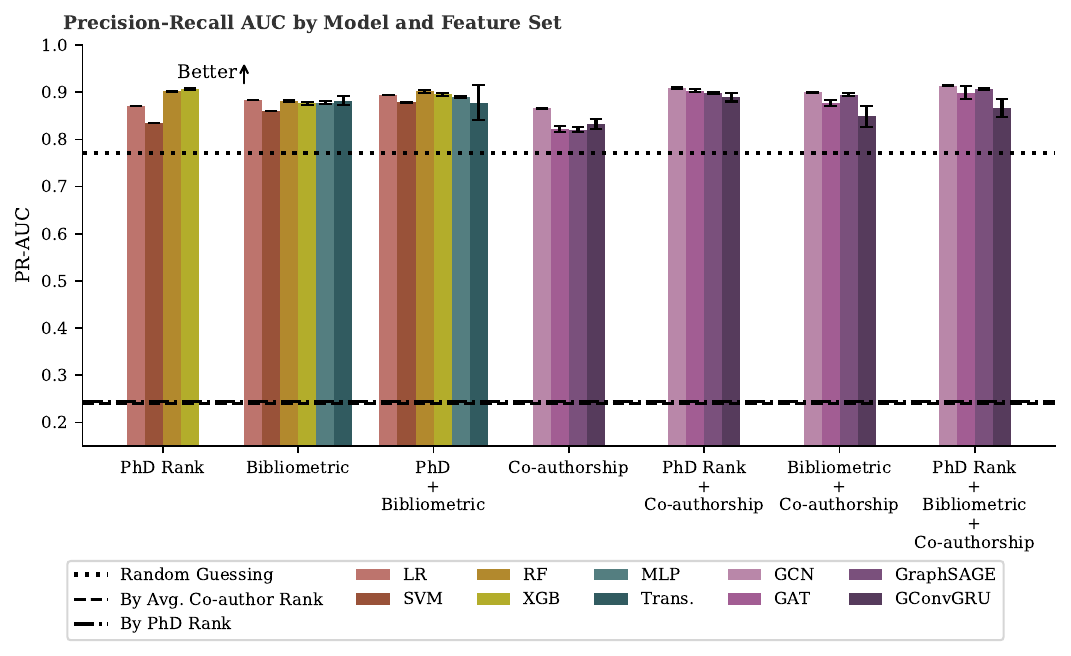}
\caption{\textbf{Predictive performance for top-$\mathbf{50}$ department placement by feature set and model type.} We plot the average PR-AUC and standard deviation across $10$ runs for each of the model trained on our feature sets. Higher values of PR-AUC indicate better performance, and we compare against our three heuristics: random guessing (dotted), the average co-author rank (dashed), and PhD rank (dash-dotted).}
\label{fig:bar_chart_y50_rewiring}
\end{figure}

\begin{figure}[t]
\centering
\includegraphics[width=.9\textwidth]{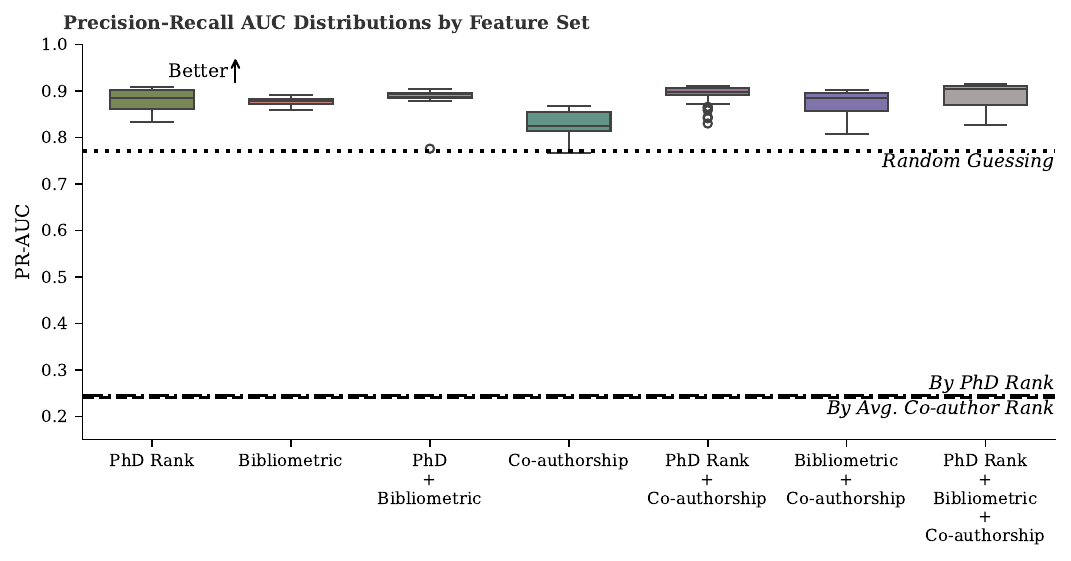}
\caption{\textbf{Predictive performance for top-$\mathbf{50}$ department placement grouped by feature set.} We plot comparative boxplots of the distribution of PR-AUC scores for the models trained on each combination of our feature sets. The performance breakdown by model can be found in Figure~\ref{fig:bar_chart_y50_rewiring}. Higher values of PR-AUC indicate better performance, and we compare against our three heuristics: random guessing (dotted), the average co-author rank (dashed), and PhD rank (dash-dotted).}
\label{fig:box_y50}
\end{figure}

Performance results for top-$30$, $40$, and $50$ hiring thresholds follow a similar structure. Figures~\ref{fig:bar_chart_y30_rewiring},~\ref{fig:bar_chart_y40_rewiring}, and~\ref{fig:bar_chart_y50_rewiring} show model-level comparisons for each feature set, while Tables~\ref{tab:best_model_y30}-\ref{tab:best_model_y50} report the best-performing model for each feature set. Box plots are provided in Figures~\ref{fig:box_y30}-\ref{fig:box_y50}.

As in the top-$10$ and top-$20$ cases, the strongest models generally incorporate all three feature types, except in the top-$30$ condition, where the best-performing model uses only PhD rank and co-authorship features (Table~\ref{tab:best_model_y30}). Across all thresholds, the average performance gap between feature sets narrows as the definition of high-rank becomes less selective. These results are summarized in Figure~\ref{fig:robustness} and discussed in Section~\ref{sec:results:robustness}. This trend is likely due in part to how the increased number of positive examples at higher thresholds might make it easier to identify high-rank features, but also reflects the behavior of PR-AUC as an evaluation metric. Since the baseline PR-AUC corresponds to the proportion of positive labels in the dataset, model performance becomes increasingly compressed as this proportion approaches $1.0$.

\section{Full Statistical Results for Broadened Definitions of \textit{High-Rank}}
\label{sec:SI:full_stats}

Appendices~\ref{sec:SI:top_10_results} and~\ref{sec:SI:full_results} provide detailed results on model performance for predicting whether researchers are hired at top-$10$, $20$, $30$, $40$, and $50$ departments. For top-$10$ placement, Table~\ref{tab:stats_y10} shows the linear mixed-effects regression results. Here, we present the full statistical results for models evaluated at the top-$20$, $30$, $40$, and $50$ thresholds in Tables~\ref{tab:stats_y20},-\ref{tab:stats_y50}.

As discussed in Section~\ref{sec:results:robustness} and shown in Figure~\ref{fig:robustness}, the contribution of co-authorship features diminishes as the high-rank threshold broadens from top-$10$ to top-$50$ departments. Meanwhile, the predictive value of PhD rank features becomes increasingly prominent. This pattern is reflected in the blue and green rows of Tables~\ref{tab:stats_y20}-\ref{tab:stats_y50}, which track the marginal gains from co-authorship and PhD rank, respectively. However, as shown in Figure~\ref{fig:robustness} and highlighted in more detail in the tables, the differences in performance between feature sets tend not to be statistically significant at higher thresholds.

%
% y_10 box
%
\FloatBarrier
%
% y_10 stats
%
\begin{table}[p!]
\centering
\resizebox{\textwidth}{!}{%
\begin{tabular}{lcccccc}
\multicolumn{7}{l}{\textbf{Mixed Linear Model Regression Summary}} \\
\\
\multicolumn{1}{l}{Model:} & \multicolumn{1}{l}{MixedLM} & \multicolumn{1}{l}{Dependent Variable:} & \multicolumn{1}{l}{PR-AUC} & \multicolumn{1}{l}{Method:} & \multicolumn{1}{l}{REML} & \\
\multicolumn{1}{l}{Scale:} & \multicolumn{1}{l}{0.0011} & \multicolumn{1}{l}{Log-Likelihood:} & \multicolumn{1}{l}{912.5417} & \multicolumn{1}{l}{Converged:} & \multicolumn{1}{l}{Yes} & \\
\multicolumn{1}{l}{No. Observations:} & \multicolumn{1}{l}{480} & \multicolumn{1}{l}{No. Groups:} & \multicolumn{1}{l}{10} & \multicolumn{1}{l}{Min. group size:} & \multicolumn{1}{l}{20} & \\
\multicolumn{1}{l}{Max. group size:} & \multicolumn{1}{l}{80} & \multicolumn{1}{l}{Mean group size:} & \multicolumn{1}{l}{48.0} &  &  & \\
& & & & & & \\
\multicolumn{7}{l}{\textbf{Reference = PhD Rank}} \\
Term & Coef. & Std. Err. & $z$ & $P>|z|$ & CI Lower & CI Upper \\
\hline
Intercept & $0.342$ & $0.007$ & $45.797$ & $0.000$ & $0.327$ & $0.357$ \\
{\color[HTML]{009c4a}Bibliometric} & {\color[HTML]{009c4a}$0.025$} & {\color[HTML]{009c4a}$0.007$} & {\color[HTML]{009c4a}$3.512$} & {\color[HTML]{009c4a}$0.000$} & {\color[HTML]{009c4a}$0.011$} & {\color[HTML]{009c4a}$0.039$} \\
Bibliometric $+$ Co-authorship & $0.062$ & $0.010$ & $6.008$ & $0.000$ & $0.042$ & $0.082$ \\
PhD Rank $+$ Bibliometric & $0.053$ & $0.007$ & $7.530$ & $0.000$ & $0.040$ & $0.067$ \\
PhD Rank $+$ Bibliometric $+$ Co-authorship & $0.082$ & $0.010$ & $8.006$ & $0.000$ & $0.062$ & $0.103$ \\
{\color[HTML]{3531FF}PhD Rank $+$ Co-authorship} & {\color[HTML]{3531FF}$0.029$} & {\color[HTML]{3531FF}$0.010$} & {\color[HTML]{3531FF}$2.828$} & {\color[HTML]{3531FF}$0.005$} & {\color[HTML]{3531FF}$0.009$} & {\color[HTML]{3531FF}$0.049$} \\
Co-authorship & $-0.069$ & $0.010$ & $-6.692$ & $0.000$ & $-0.089$ & $-0.049$ \\
Group Variance & $0.000$ & $0.003$ &  &  & &  \\
& & & & & & \\
\multicolumn{7}{l}{\textbf{Reference = Bibliometric}} \\
Term & Coef. & Std. Err. & $z$ & $P>|z|$ & CI Lower & CI Upper \\
\hline
Intercept & $0.367$ & $0.007$ & $56.021$ & $0.000$ & $0.354$ & $0.380$ \\
{\color[HTML]{3531FF}Bibliometric $+$ Co-authorship} & {\color[HTML]{3531FF}$0.037$} & {\color[HTML]{3531FF}$0.010$} & {\color[HTML]{3531FF}$3.827$} & {\color[HTML]{3531FF}$0.000$} & {\color[HTML]{3531FF}$0.018$} & {\color[HTML]{3531FF}$0.056$} \\
PhD Rank & $-0.025$ & $0.007$ & $-3.512$ & $0.000$ & $-0.039$ & $-0.011$ \\
PhD Rank $+$ Bibliometric & $0.029$ & $0.006$ & $4.656$ & $0.000$ & $0.017$ & $0.041$ \\
PhD Rank $+$ Bibliometric $+$ Co-authorship & $0.058$ & $0.010$ & $5.959$ & $0.000$ & $0.039$ & $0.076$ \\
PhD Rank $+$ Co-authorship & $0.004$ & $0.010$ & $0.435$ & $0.663$ & $-0.015$ & $0.023$ \\
Co-authorship & $-0.094$ & $0.010$ & $-9.721$ & $0.000$ & $-0.113$ & $-0.075$ \\
Group Variance & $0.000$ & $0.003$ &  &  & &  \\
& & & & & & \\
\multicolumn{7}{l}{\textbf{Reference = PhD Rank $+$ Bibliometric}} \\
Term & Coef. & Std. Err. & $z$ & $P>|z|$ & CI Lower & CI Upper \\
\hline
Intercept & $0.396$ & $0.007$ & $60.374$ & $0.000$ & $0.383$ & $0.408$ \\
Bibliometric & $-0.029$ & $0.006$ & $-4.656$ & $0.000$ & $-0.041$ & $-0.017$ \\
Bibliometric $+$ Co-authorship & $0.008$ & $0.010$ & $0.874$ & $0.382$ & $-0.010$ & $0.027$ \\
PhD Rank & $-0.053$ & $0.007$ & $-7.530$ & $0.000$ & $-0.067$ & $-0.040$ \\
{\color[HTML]{3531FF}PhD Rank $+$ Bibliometric $+$ Co-authorship} & {\color[HTML]{3531FF}$0.029$} & {\color[HTML]{3531FF}$0.010$} & {\color[HTML]{3531FF}$3.005$} & {\color[HTML]{3531FF}$0.003$} & {\color[HTML]{3531FF}$0.010$} & {\color[HTML]{3531FF}$0.048$} \\
PhD Rank $+$ Co-authorship & $-0.024$ & $0.010$ & $-2.518$ & $0.012$ & $-0.043$ & $-0.005$ \\
Co-authorship & $-0.122$ & $0.010$ & $-12.675$ & $0.000$ & $-0.141$ & $-0.103$ \\
Group Variance & $0.000$ & $0.003$ &  &  & &  \\
& & & & & & \\
\multicolumn{7}{l}{\textbf{Reference = PhD Rank $+$ Co-authorship}} \\
Term & Coef. & Std. Err. & $z$ & $P>|z|$ & CI Lower & CI Upper \\
\hline
Intercept & $0.371$ & $0.007$ & $52.335$ & $0.000$ & $0.357$ & $0.385$ \\
Bibliometric & $-0.004$ & $0.010$ & $-0.453$ & $0.663$ & $-0.023$ & $0.015$ \\
{\color[HTML]{009c4a}Bibliometric $+$ Co-authorship} & {\color[HTML]{009c4a}$0.033$} & {\color[HTML]{009c4a}$0.005$} & {\color[HTML]{009c4a}$6.175$} & {\color[HTML]{009c4a}$0.000$} & {\color[HTML]{009c4a}$0.022$} & {\color[HTML]{009c4a}$0.043$} \\
PhD Rank & $-0.029$ & $0.010$ & $-2.828$ & $0.005$ & $-0.049$ & $-0.009$ \\
PhD Rank $+$ Bibliometric & $0.024$ & $0.010$ & $2.518$ & $0.012$ & $0.005$ & $0.043$\\
PhD Rank $+$ Bibliometric $+$ Co-authorship & $0.053$ & $0.005$ & $10.054$ & $0.000$ & $0.043$ & $0.064$ \\
Co-authorship & $-0.098$ & $0.005$ & $-18.487$ & $0.000$ & $-0.108$ & $-0.088$ \\
Group Variance & $0.000$ & $0.003$ &  &  & &  \\
\end{tabular}
}
\caption{\textbf{Linear mixed-effects modeling results for predicting faculty placement at top-$\mathbf{10}$ departments.} A summary of the linear mixed-effects regression model used to predict the PR-AUC of each model-feature combination as a function fo the feature set (fixed effect) and model architecture (random effect). We provide information relating to running the model including total number of observations and groups, the minimum, maximum, and mean group size, and the method. We consider four different reference feature sets (i.e., intercepts): (1) PhD rank, (2) bibliometric features, (3) PhD rank and bibliometric features, and (4) PhD rank and the co-authorhship network. Rows in blue highlight the feature sets which add the co-authorship network to the reference features and are used to generate Figure~\ref{fig:robustness}(a), while rows in green compare the PhD rank and bibliometric feature sets and are used to generate Figure~\ref{fig:robustness}(b).}
\label{tab:stats_y10}
\end{table}
%
% y_20 table
%
\begin{table}[htb]
\centering
\resizebox{\textwidth}{!}{%
\begin{tabular}{c|c|c|l|c|c|c|c|c}
\textbf{PhD} & \textbf{Bib} & \textbf{Co-author} & \textbf{Optimal Model} & \textbf{\begin{tabular}[c]{@{}c@{}}Precision\\ (std)\end{tabular}} & \textbf{\begin{tabular}[c]{@{}c@{}}Recall\\ (std)\end{tabular}} & \textbf{\begin{tabular}[c]{@{}c@{}}PR-AUC\\ (std)\end{tabular}} & \textbf{\begin{tabular}[c]{@{}c@{}}\% Improvement \\ over PhD Rank\end{tabular}} & \textbf{\begin{tabular}[c]{@{}c@{}}\% Improvement over \\ Avg. Co-author Rank\end{tabular}} \\ \hline \hline
    &                       &                        & Random             &-                                                                  & -                                                               & $0.357$                                                         & $17.43\%$                                                                        & $25.70\%$                                                                                  \\
 \checkmark                   &                       &                        & PhD Rank           &$0.322$ $(0.000)$                                                  & $\mathbf{0.822\,  (0.000)}$                                      & $0.304$ $(0.000)$                                               & $0.00\%$                                                                         & $7.04\%$                                                                                   \\
   &                       & \checkmark             & Avg. Co-author Rank & $0.340$ $(0.000)$                                                  & $0.527$ $(0.000)$                                               & $0.284$ $(0.000)$                                               & $-6.58\%$                                                                        & $0.00\%$                                                                                   \\ \hline
\checkmark                   &                       &                        & RF                 &  $0.537$ $(0.005)$                                                  & $0.628$ $(0.010)$                                               & $0.564$ $(0.008)$                        & $85.53\%$                                                 & $98.59\%$                                                           \\
      & \checkmark            &                        & LR                 & $0.519$ $(0.000)$                                                  & $0.654$ $(0.000)$                                               & $0.563$ $(0.000)$                        &  $85.20\%$                                                 & $98.24\%$                                                           \\
     &                       & \checkmark             & GConvGRU           &  $0.495$ $(0.050)$                                                  & $0.522$ $(0.182)$                                               & $0.503$ $(0.037)$                                               & $65.46\%$                                                                        & $77.11\%$                                                                                  \\
\checkmark                   & \checkmark            &                        & SVC                &  $0.544$ $(0.000)$                                                  & $0.563$ $(0.000)$                                               & $0.588$ $(0.000)$                                               & $93.42\%$                                                                        & $107.04\%$                                                                                 \\
 \checkmark                   &                       & \checkmark             & GConvGRU           & $\mathbf{0.546\, (0.039)}$                                         & $0.747$ $(0.072)$                                               &  $0.575$ $(0.044)$                        &  $89.14\%$                                                 & $102.46\%$                                                          \\
     & \checkmark            & \checkmark             & GCN                & $0.484$ $(0.005)$                                                  & $0.610$ $(0.012)$                                               &  $0.564$ $(0.005)$                        & $85.53\%$                                               & $98.59\%$                                                           \\
 \checkmark                   & \checkmark            & \checkmark             & GraphSAGE          & $0.537$ $(0.012)$                                                  & $0.655$ $(0.020)$                                               & {\color{blue} $0.589$ $(0.009)$}                        & {\color{blue} $93.75\%$}                                                 & {\color{blue} $107.39\%$}                                                     
\end{tabular}
}
\caption{\textbf{Average performance of the best model for each feature set when predicting hire at top-$\mathbf{20}$ departments.} We identify the top-performing model for each feature set based on the average PR-AUC score over $10$ runs. We break down the PR-AUC score into its components, precision and recall, and calculate the percent improvement over the PhD rank and average co-author rank heuristics. Precision and recall are computed using a threshold of $0.5$ on the predicted class probabilities. The proportion of faculty hired at top-$10$ departments is equivalent to the PR-AUC of the random guessing baseline (i.e., $0.357$). The best-performing model by PR-AUC is listed in blue, while the best precision and recall are bolded.}
\label{tab:best_model_y20}
\end{table}
%
% y_20 bar chart
%
%
% y_20 box
%
%
%
% y_30 table
%
\begin{table}[t]
\centering
\resizebox{\textwidth}{!}{%
\begin{tabular}{c|c|c|l|c|c|c|c|c}
\textbf{PhD} & \textbf{Bib} & \textbf{Co-author} & \textbf{Optimal Model}     & \textbf{\begin{tabular}[c]{@{}c@{}}Precision\\ (std)\end{tabular}} & \textbf{\begin{tabular}[c]{@{}c@{}}Recall\\ (std)\end{tabular}} & \textbf{\begin{tabular}[c]{@{}c@{}}PR-AUC\\ (std)\end{tabular}} & \textbf{\begin{tabular}[c]{@{}c@{}}\% Improvement \\ over PhD Rank\end{tabular}} & \textbf{\begin{tabular}[c]{@{}c@{}}\% Improvement over \\ Avg. Co-author Rank\end{tabular}} \\ \hline \hline
&                       &                        & Random             &                   -                                                                  & -                                                               & $0.491$                                                         & $85.98\%$                                                                        & $84.59\%$                                                                                  \\
\checkmark        &                       &                        & PhD Rank           & $0.266$ $(0.000)$                                                  & $\mathbf{0.961\, (0.000)}$                                      & $0.264$ $(0.000)$                                               & $0.00\%$                                                                         & $-0.75\%$                                                                                  \\
&                       & \checkmark             & Avg. Co-author Rank &                   $0.283$ $(0.000)$                                                  & $0.729$ $(0.000)$                                               & $0.266$ $(0.000)$                                               & $0.76\%$                                                                         & $0.00\%$                                                                                   \\ \hline
\checkmark        &                       &                        & RF                 & $0.673$ $(0.001)$                                                  & $0.587$ $(0.010)$                                               & $0.688$ $(0.003)$                       &  $160.61\%$                                                & $158.65\%$                                                          \\
& \checkmark            &                        & XGB                &                   $0.653$ $(0.004)$                                                  & $0.594$ $(0.007)$                                               & $0.662$ $(0.005)$                        & $150.76\%$                                                & $148.87\%$                                                          \\
 &                       & \checkmark             & GCN                &                  $0.584$ $(0.011)$                                                  & $0.479$ $(0.025)$                                               & $0.611$ $(0.009)$                                               & $131.44\%$                                                                       & $129.70\%$                                                                                 \\
\checkmark        & \checkmark            &                        & LR                 & $0.640$ $(0.000)$                                                  & $0.759$ $(0.000)$                                               & $0.701$ $(0.000)$                                               & $165.53\%$                                                                       & $163.53\%$                                                                                 \\
\checkmark        &                       & \checkmark             & GConvGRU           & $0.663$ $(0.041)$                                                  & $0.721$ $(0.100)$                                               & {\color{blue} $0.712$ $(0.039)$}                        & {\color{blue} $169.70\%$}                                                & {\color{blue} $167.67\%$}                                                          \\
& \checkmark            & \checkmark             & GCN                &                   $\mathbf{0.680\, (0.007)}$                                         & $0.573$ $(0.008)$                                               & $0.698$ $(0.002)$                        & $164.39\%$                                                & $162.41\%$                                                          \\
\checkmark        & \checkmark            & \checkmark             & GCN                & $0.667$ $(0.004)$                                                  & $0.646$ $(0.007)$                                               & $0.695$ $(0.005)$                                               & $163.26\%$                                                                       & $161.28\%$                                                                                
\end{tabular}
}
\caption{\textbf{Average performance of the best model for each feature set when predicting hire at top-$\mathbf{30}$ departments.} We identify the top-performing model for each feature set based on the average PR-AUC score over $10$ runs. We break down the PR-AUC score into its components, precision and recall, and calculate the percent improvement over the PhD rank and average co-author rank heuristics. Precision and recall are computed using a threshold of $0.5$ on the predicted class probabilities. The proportion of faculty hired at top-$10$ departments is equivalent to the PR-AUC of the random guessing baseline (i.e., $0.491$). The best-performing model by PR-AUC is listed in blue, while the best precision and recall are bolded.}
\label{tab:best_model_y30}
\end{table}
%
% y_30 bar chart
%
%
% y_30 box
%
%
%
% y_40 table
%
\begin{table}[t]
\centering
\resizebox{\textwidth}{!}{%
\begin{tabular}{c|c|c|l|c|c|c|c|c}
\textbf{PhD} & \textbf{Bib} & \textbf{Co-author} & \textbf{Optimal Model}     & \textbf{\begin{tabular}[c]{@{}c@{}}Precision\\ (std)\end{tabular}} & \textbf{\begin{tabular}[c]{@{}c@{}}Recall\\ (std)\end{tabular}} & \textbf{\begin{tabular}[c]{@{}c@{}}PR-AUC\\ (std)\end{tabular}} & \textbf{\begin{tabular}[c]{@{}c@{}}\% Improvement \\ over PhD Rank\end{tabular}} & \textbf{\begin{tabular}[c]{@{}c@{}}\% Improvement over \\ Avg. Co-author Rank\end{tabular}} \\ \hline \hline
&                       &                        & Random             &                   -                                                                  & -                                                               & $0.674$                                                         & $166.40\%$                                                                       & $179.67\%$                                                                                 \\
\checkmark        &                       &                        & PhD Rank           & $0.254$ $(0.000)$                                                  & $0.969$ $(0.000)$                                               & $0.253$ $(0.000)$                                               & $0.00\%$                                                                         & $4.98\%$                                                                                   \\
&                       & \checkmark             & Avg. Co-author Rank &                   $0.246$ $(0.000)$                                                  & $0.806$ $(0.000)$                                               & $0.241$ $(0.000)$                                               & $-4.74\%$                                                                        & $0.00\%$                                                                                   \\ \hline
\checkmark        &                       &                        & XGB                & $0.711$ $(0.001)$                                                  & $0.975$ $(0.017)$                                               & $0.839$ $(0.001)$                        &  $231.62\%$                                                &  $248.13\%$                                                          \\
& \checkmark            &                        & LR                 &                   $0.710$ $(0.000)$                                                  & $\mathbf{0.969 (0.000)}$                                        & $0.832$ $(0.000)$                        &  $228.85\%$                                                & $245.23\%$                                                          \\
&                       & \checkmark             & GCN                &                   $0.744$ $(0.009)$                                                  & $0.496$ $(0.019)$                                               & $0.765$ $(0.003)$                                               & $202.37\%$                                                                       & $217.43\%$                                                                                 \\
\checkmark        & \checkmark            &                        & LR                 & $0.744$ $(0.000)$                                                  & $0.934$ $(0.000)$                                               & $0.845$ $(0.000)$                                               & $233.99\%$                                                                       & $250.62\%$                                                                                 \\
\checkmark        &                       & \checkmark             & GCN                & $\mathbf{0.795\, (0.011)}$                                         & $0.685$ $(0.033)$                                               & $0.822$ $(0.003)$                                               & $224.90\%$                                                                       & $241.08\%$                                                                                 \\
& \checkmark            & \checkmark             & GCN                &                   $0.790$ $(0.003)$                                                  & $0.613$ $(0.010)$                                               & $0.843$ $(0.002)$                                               & $233.20\%$                                                                       & $249.79\%$                                                                                 \\
\checkmark        & \checkmark            & \checkmark             & GCN                & $0.788$ $(0.004)$                                                  & $0.717$ $(0.005)$                                               & {\color{blue} $0.846$ $(0.003)$}                        & {\color{blue} $234.39\%$}                                                & {\color{blue} $251.04\%$}                                                         
\end{tabular}
}
\caption{\textbf{Average performance of the best model for each feature set when predicting hire at top-$\mathbf{40}$ departments.} We identify the top-performing model for each feature set based on the average PR-AUC score over $10$ runs. We break down the PR-AUC score into its components, precision and recall, and calculate the percent improvement over the PhD rank and average co-author rank heuristics. Precision and recall are computed using a threshold of $0.5$ on the predicted class probabilities. The proportion of faculty hired at top-$10$ departments is equivalent to the PR-AUC of the random guessing baseline (i.e., $0.674$). The best-performing model by PR-AUC is listed in blue, while the best precision and recall are bolded.}
\label{tab:best_model_y40}
\end{table}
%
% y_40 bar chart
%
%
% y_40 box
%
%
%
% y_50 table
%
\begin{table}[t]
\centering
\resizebox{\textwidth}{!}{%
\begin{tabular}{c|c|c|l|c|c|c|c|c}
\textbf{PhD} & \textbf{Bib} & \textbf{Co-author} & \textbf{Optimal Model}     & \textbf{\begin{tabular}[c]{@{}c@{}}Precision\\ (std)\end{tabular}} & \textbf{\begin{tabular}[c]{@{}c@{}}Recall\\ (std)\end{tabular}} & \textbf{\begin{tabular}[c]{@{}c@{}}PR-AUC\\ (std)\end{tabular}} & \textbf{\begin{tabular}[c]{@{}c@{}}\% Improvement \\ over PhD Rank\end{tabular}} & \textbf{\begin{tabular}[c]{@{}c@{}}\% Improvement over \\ Avg. Co-author Rank\end{tabular}} \\ \hline \hline
&                       &                        & -                                                                  & -                                                               & Random             &                   $0.771$                                                         & $214.69\%$                                                                       & $221.25\%$                                                                                 \\
\checkmark        &                       &                        & PhD Rank           & $0.245$ $(0.000)$                                                  & $0.977$ $(0.000)$                                               & $0.245$ $(0.000)$                                               & $0.00\%$                                                                         & $2.08\%$                                                                                   \\
 &                       & \checkmark             & Avg. Co-author Rank &                  $0.243$ $(0.000)$                                                  & $0.876$ $(0.000)$                                               & $0.240$ $(0.000)$                                               & $-2.04\%$                                                                        & $0.00\%$                                                                                   \\ \hline
\checkmark        &                       &                        & XGB                & $0.812$ $(0.005)$                                                  & $0.972$ $(0.004)$                                               &  $0.907$ $(0.002)$                        &  $270.20\%$                                                &  $277.92\%$                                                          \\
& \checkmark            &                        & LR                 &                   $0.797$ $(0.000)$                                                  & $\mathbf{0.980\, (0.000)}$                                      &  $0.883$ $(0.000)$                        &  $260.41\%$                                                &  $267.92\%$                                                          \\
&                       & \checkmark             & GCN                &                   $0.849$ $(0.005)$                                                  & $0.503$ $(0.015)$                                               & $0.865$ $(0.001)$                                               & $253.06\%$                                                                       & $260.42\%$                                                                                 \\
\checkmark        & \checkmark            &                        & RF                 & $0.818$ $(0.003)$                                                  & $0.966$ $(0.005)$                                               & $0.902$ $(0.003)$                                               & $268.16\%$                                                                       & $275.83\%$                                                                                 \\
\checkmark        &                       & \checkmark             & GCN                & $\mathbf{0.900\, (0.010)}$                                         & $0.620$ $(0.030)$                                               & $0.908$ $(0.002)$                        &  $270.61\%$                                                &  $278.33\%$                                                         \\
& \checkmark            & \checkmark             & GCN                &                   $0.860$ $(0.003)$                                                  & $0.672$ $(0.011)$                                               &  $0.899$ $(0.002)$                        &  $266.94\%$                                                & $274.58\%$                                                          \\
\checkmark        & \checkmark            & \checkmark             & GCN                & $0.852$ $(0.010)$                                                  & $0.789$ $(0.023)$                                               & {\color{blue} $0.914$ $(0.001)$}                        & {\color{blue} $273.06\%$}                                                & {\color{blue} $280.83\%$}                                                         
\end{tabular}
}
\caption{\textbf{Average performance of the best model for each feature set when predicting hire at top-$\mathbf{50}$ departments.} We identify the top-performing model for each feature set based on the average PR-AUC score over $10$ runs. We break down the PR-AUC score into its components, precision and recall, and calculate the percent improvement over the PhD rank and average co-author rank heuristics. Precision and recall are computed using a threshold of $0.5$ on the predicted class probabilities. The proportion of faculty hired at top-$10$ departments is equivalent to the PR-AUC of the random guessing baseline (i.e., $0.771$). The best-performing model by PR-AUC is listed in blue, while the best precision and recall are bolded.}
\label{tab:best_model_y50}
\end{table}
%
% y_50 bar chart
%
%
% y_50 box
%
%
% y_20 stats no rewiring
%
\begin{table}[t]
\centering
\resizebox{\textwidth}{!}{%
\begin{tabular}{lcccccc}
\multicolumn{7}{l}{\textbf{Mixed Linear Model Regression Summary}} \\
\\
\multicolumn{1}{l}{Model:} & \multicolumn{1}{l}{MixedLM} & \multicolumn{1}{l}{Dependent Variable:} & \multicolumn{1}{l}{PR-AUC} & \multicolumn{1}{l}{Method:} & \multicolumn{1}{l}{REML} & \\
\multicolumn{1}{l}{Scale:} & \multicolumn{1}{l}{0.0014} & \multicolumn{1}{l}{Log-Likelihood:} & \multicolumn{1}{l}{867.3958} & \multicolumn{1}{l}{Converged:} & \multicolumn{1}{l}{Yes} & \\
\multicolumn{1}{l}{No. Observations:} & \multicolumn{1}{l}{480} & \multicolumn{1}{l}{No. Groups:} & \multicolumn{1}{l}{10} & \multicolumn{1}{l}{Min. group size:} & \multicolumn{1}{l}{20} & \\
\multicolumn{1}{l}{Max. group size:} & \multicolumn{1}{l}{80} & \multicolumn{1}{l}{Mean group size:} & \multicolumn{1}{l}{48.0} &  &  & \\
& & & & & & \\
\multicolumn{7}{l}{\textbf{Reference = PhD Rank}} \\
Term & Coef. & Std. Err. & $z$ & $P>|z|$ & CI Lower & CI Upper \\
\hline
Intercept & $0.536$ & $0.008$ & $65.473$ & $0.000$ & $0.520$ & $0.552$ \\
{\color[HTML]{009c4a}Bibliometric} & {\color[HTML]{009c4a}$-0.006$} & {\color[HTML]{009c4a}$0.008$} & {\color[HTML]{009c4a}$-0.712$} & {\color[HTML]{009c4a}$0.476$} & {\color[HTML]{009c4a}$-0.021$} & {\color[HTML]{009c4a}$0.010$} \\
Bibliometric $+$ Co-authorship & $0.003$ & $0.011$ & $0.271$ & $0.787$ & $-0.019$ & $0.025$ \\
PhD Rank $+$ Bibliometric & $0.029$ & $0.008$ & $3.667$ & $0.000$ & $0.013$ & $0.044$ \\
PhD Rank $+$ Bibliometric $+$ Co-authorship & $0.025$ & $0.011$ & $2.241$ & $0.025$ & $0.003$ & $0.047$ \\
{\color[HTML]{3531FF}PhD Rank $+$ Co-authorship} & {\color[HTML]{3531FF}$0.006$} & {\color[HTML]{3531FF}$0.011$} & {\color[HTML]{3531FF}$0.526$} & {\color[HTML]{3531FF}$0.599$} & {\color[HTML]{3531FF}$-0.016$} & {\color[HTML]{3531FF}$0.028$} \\
Co-authorship & $-0.102$ & $0.011$ & $-9.004$ & $0.000$ & $-0.124$ & $-0.079$ \\
Group Variance & $0.000$ & $0.003$ &  &  & &  \\
& & & & & & \\
\multicolumn{7}{l}{\textbf{Reference = Bibliometric}} \\
Term & Coef. & Std. Err. & $z$ & $P>|z|$ & CI Lower & CI Upper \\
\hline
Intercept & $0.530$ & $0.007$ & $73.855$ & $0.000$ & $0.516$ & $0.545$ \\
{\color[HTML]{3531FF}Bibliometric $+$ Co-authorship} & {\color[HTML]{3531FF}$0.009$} & {\color[HTML]{3531FF}$0.011$} & {\color[HTML]{3531FF}$0.814$} &{\color[HTML]{3531FF} $0.416$} & {\color[HTML]{3531FF}$-0.012$} & {\color[HTML]{3531FF}$0.029$} \\
PhD Rank & $0.006$ & $0.008$ & $0.712$ & $0.476$ & $-0.010$ & $0.021$ \\
PhD Rank $+$ Bibliometric & $0.034$ & $0.007$ & $5.069$ & $0.000$ & $0.021$ & $0.047$ \\
PhD Rank $+$ Bibliometric $+$ Co-authorship & $0.031$ & $0.011$ & $2.916$ & $0.004$ & $0.010$ & $0.052$ \\
PhD Rank $+$ Co-authorship & $0.011$ & $0.011$ & $1.086$ & $0.277$ & $-0.009$ & $0.032$ \\
Co-authorship & $-0.096$ & $0.011$ & $-9.079$ & $0.000$ & $-0.117$ & $-0.075$ \\
Group Variance & $0.000$ & $0.003$ &  &  & &  \\
& & & & & & \\
\multicolumn{7}{l}{\textbf{Reference = PhD Rank $+$ Bibliometric}} \\
Term & Coef. & Std. Err. & $z$ & $P>|z|$ & CI Lower & CI Upper \\
\hline
Intercept & $0.565$ & $0.007$ & $78.612$ & $0.000$ & $0.551$ & $0.579$ \\
Bibliometric & $-0.034$ & $0.007$ & $-5.069$ & $0.000$ & $-0.047$ & $-0.021$ \\
Bibliometric $+$ Co-authorship & $-0.026$ & $0.011$ & $-2.415$ & $0.016$ & $-0.046$ & $-0.005$ \\
PhD Rank & $-0.029$ & $0.008$ & $-3.667$ & $0.000$ & $-0.044$ & $-0.013$ \\
{\color[HTML]{3531FF}PhD Rank $+$ Bibliometric $+$ Co-author} & {\color[HTML]{3531FF}$-0.003$} & {\color[HTML]{3531FF}$0.011$} & {\color[HTML]{3531FF}$-0.314$} & {\color[HTML]{3531FF}$0.754$} & {\color[HTML]{3531FF}$-0.024$} & {\color[HTML]{3531FF}$0.017$} \\
PhD Rank $+$ Co-authorship & $-0.023$ & $0.011$ & $-2.143$ & $0.032$ & $-0.043$ & $-0.002$ \\
Co-authorship & $-0.130$ & $0.011$ & $-12.309$ & $0.000$ & $-0.151$ & $-0.109$ \\
Group Variance & $0.000$ & $0.003$ &  &  & &  \\
& & & & & & \\
\multicolumn{7}{l}{\textbf{Reference = PhD Rank $+$ Co-authorship}} \\
Term & Coef. & Std. Err. & $z$ & $P>|z|$ & CI Lower & CI Upper \\
\hline
Intercept & $0.542$ & $0.008$ & $69.765$ & $0.000$ & $0.527$ & $0.557$ \\
Bibliometric & $-0.011$ & $0.011$ & $-1.086$ & $0.277$ & $-0.032$ & $0.009$ \\
{\color[HTML]{009c4a}Bibliometric $+$ Co-authorship} & {\color[HTML]{009c4a}$-0.003$} & {\color[HTML]{009c4a}$0.006$} & {\color[HTML]{009c4a}$-0.494$} & {\color[HTML]{009c4a}$0.621$} & {\color[HTML]{009c4a}$-0.014$} & {\color[HTML]{009c4a}$0.009$} \\
PhD Rank & $-0.006$ & $0.011$ & $-0.526$ & $0.599$ & $-0.028$ & $0.016$ \\
PhD Rank $+$ Bibliometric & $0.023$ & $0.011$ & $2.143$ & $0.032$ & $0.002$ & $0.043$ \\
PhD Rank $+$ Bibliometric $+$ Co-authorship & $0.019$ & $0.006$ & $3.315$ & $0.001$ & $0.008$ & $0.031$ \\
Co-authorship & $-0.108$ & $0.006$ & $-18.426$ & $0.000$ & $-0.119$ & $-0.096$ \\
Group Variance & $0.000$ & $0.003$ &  &  & &  \\
\end{tabular}
}
\caption{\textbf{Linear mixed-effects modeling results for predicting faculty placement at top-$\mathbf{20}$ departments.} A summary of the linear mixed-effects regression model used to predict the PR-AUC of each model-feature combination as a function fo the feature set (fixed effect) and model architecture (random effect). We provide information relating to running the model including total number of observations and groups, the minimum, maximum, and mean group size, and the method. We consider four different reference feature sets (i.e., intercepts): (1) PhD rank, (2) bibliometric features, (3) PhD rank and bibliometric features, and (4) PhD rank and the co-authorhship network. Rows in blue highlight the feature sets which add the co-authorship network to the reference features and are used to generate Figure~\ref{fig:robustness}(a), while rows in green compare the PhD rank and bibliometric feature sets and are used to generate Figure~\ref{fig:robustness}(b).}
\label{tab:stats_y20}
\end{table}
%
% y_30 stats no rewiring
%
\begin{table}[t]
\centering
\resizebox{\textwidth}{!}{%
\begin{tabular}{lcccccc}
\multicolumn{7}{l}{\textbf{Mixed Linear Model Regression Summary}} \\
\\
\multicolumn{1}{l}{Model:} & \multicolumn{1}{l}{MixedLM} & \multicolumn{1}{l}{Dependent Variable:} & \multicolumn{1}{l}{PR-AUC} & \multicolumn{1}{l}{Method:} & \multicolumn{1}{l}{REML} & \\
\multicolumn{1}{l}{Scale:} & \multicolumn{1}{l}{0.0012} & \multicolumn{1}{l}{Log-Likelihood:} & \multicolumn{1}{l}{893.8748} & \multicolumn{1}{l}{Converged:} & \multicolumn{1}{l}{Yes} & \\
\multicolumn{1}{l}{No. Observations:} & \multicolumn{1}{l}{480} & \multicolumn{1}{l}{No. Groups:} & \multicolumn{1}{l}{10} & \multicolumn{1}{l}{Min. group size:} & \multicolumn{1}{l}{20} & \\
\multicolumn{1}{l}{Max. group size:} & \multicolumn{1}{l}{80} & \multicolumn{1}{l}{Mean group size:} & \multicolumn{1}{l}{48.0} &  &  & \\
& & & & & & \\
\multicolumn{7}{l}{\textbf{Reference = PhD Rank}} \\
Term & Coef. & Std. Err. & $z$ & $P>|z|$ & CI Lower & CI Upper \\
\hline
Intercept & $0.666$ & $0.014$ & $47.401$ & $0.000$ & $0.639$ & $0.694$ \\
{\color[HTML]{009c4a}Bibliometric} & {\color[HTML]{009c4a}$-0.066$} & {\color[HTML]{009c4a}$0.007$} & {\color[HTML]{009c4a}$-8.962$} & {\color[HTML]{009c4a}$0.000$} & {\color[HTML]{009c4a}$-0.080$} & {\color[HTML]{009c4a}$-0.051$} \\
Bibliometric $+$ Co-authorship & $-0.008$ & $0.021$ & $-0.369$ & $0.712$ & $-0.050$ & $0.034$ \\
PhD Rank $+$ Bibliometric & $0.029$ & $0.007$ & $3.976$ & $0.000$ & $0.015$ & $0.044$ \\
PhD Rank $+$ Bibliometric $+$ Co-authorship & $0.008$ & $0.021$ & $0.381$ & $0.704$ & $-0.034$ & $0.050$ \\
{\color[HTML]{3531FF}PhD Rank $+$ Co-authorship} & {\color[HTML]{3531FF}$0.017$} & {\color[HTML]{3531FF}$0.021$} & {\color[HTML]{3531FF}$0.792$} & {\color[HTML]{3531FF}$0.429$} & {\color[HTML]{3531FF}$-0.025$} & {\color[HTML]{3531FF}$0.059$} \\
Co-authorship & $-0.107$ & $0.021$ & $-4.997$ & $0.000$ & $-0.149$ & $-0.065$ \\
Group Variance & $0.001$ & $0.015$ &  &  & & \\
& & & & & & \\
\multicolumn{7}{l}{\textbf{Reference = Bibliometric}} \\
Term & Coef. & Std. Err. & $z$ & $P>|z|$ & CI Lower & CI Upper \\
\hline
Intercept & $0.600$ & $0.014$ & $44.387$ & $0.000$ & $0.574$ & $0.627$ \\
{\color[HTML]{3531FF}Bibliometric $+$ Co-authorship} & {\color[HTML]{3531FF}$0.058$} & {\color[HTML]{3531FF}$0.021$} & {\color[HTML]{3531FF}$2.757$} & {\color[HTML]{3531FF}$0.006$} & {\color[HTML]{3531FF}$0.017$} & {\color[HTML]{3531FF}$0.099$} \\
PhD Rank & $0.066$ & $0.007$ & $8.962$ & $0.000$ & $0.051$ & $0.080$ \\
PhD Rank $+$ Bibliometric & $0.095$ & $0.006$ & $15.147$ & $0.000$ & $0.083$ & $0.107$ \\
PhD Rank $+$ Bibliometric $+$ Co-authorship & $0.074$ & $0.021$ & $3.519$ & $0.000$ & $0.033$ & $0.115$ \\
PhD Rank $+$ Co-authorship & $0.083$ & $0.021$ & $3.937$ & $0.000$ & $0.042$ & $0.124$ \\
Co-authorship & $-0.041$ & $0.021$ & $-1.946$ & $0.052$ & $-0.082$ & $0.000$ \\
Group Variance & $0.001$ & $0.015$ &  &  & & \\
& & & & & & \\
\multicolumn{7}{l}{\textbf{Reference = PhD Rank$+$ Bibliometric}} \\
Term & Coef. & Std. Err. & $z$ & $P>|z|$ & CI Lower & CI Upper \\
\hline
Intercept & $0.696$ & $0.014$ & $51.420$ & $0.000$ & $0.669$ & $0.722$ \\
Bibliometric & $-0.095$ & $0.006$ & $-15.147$ & $0.000$ & $-0.107$ & $-0.083$ \\
Bibliometric $+$ Co-authorship & $-0.037$ & $0.021$ & $-1.765$ & $0.078$ & $-0.078$ & $0.004$ \\
PhD Rank & $-0.029$ & $0.007$ & $-3.976$ & $0.000$ & $-0.044$ & $-0.015$ \\
{\color[HTML]{3531FF}PhD Rank $+$ Bibliometric $+$ Co-authorship} & {\color[HTML]{3531FF}$-0.021$} & {\color[HTML]{3531FF}$0.021$} & {\color[HTML]{3531FF}$-1.003$} & {\color[HTML]{3531FF}$0.316$} & {\color[HTML]{3531FF}$-0.062$} & {\color[HTML]{3531FF}$0.020$} \\
PhD Rank $+$ Co-authorship & $-0.012$ & $0.021$ & $-0.585$ & $0.559$ & $-0.054$ & $0.029$ \\
Co-authorship & $-0.136$ & $0.021$ & $-6.468$ & $0.000$ & $-0.177$ & $-0.095$ \\
Group Variance & $0.001$ & $0.015$ &  &  & & \\
& & & & & & \\
\multicolumn{7}{l}{\textbf{Reference = PhD Rank $+$ Co-authorship}} \\
Term & Coef. & Std. Err. & $z$ & $P>|z|$ & CI Lower & CI Upper \\
\hline
Intercept & $0.683$ & $0.016$ & $42.400$ & $0.000$ & $0.652$ & $0.715$ \\
Bibliometric & $-0.083$ & $0.021$ & $-3.937$ & $0.000$ & $-0.124$ & $-0.042$ \\
{\color[HTML]{009c4a}Bibliometric $+$ Co-authorship} & {\color[HTML]{009c4a}$-0.025$} & {\color[HTML]{009c4a}$0.005$} & {\color[HTML]{009c4a}$-4.565$} & {\color[HTML]{009c4a}$0.000$} & {\color[HTML]{009c4a}$-0.035$} & {\color[HTML]{009c4a}$-0.014$} \\
PhD Rank & $-0.017$ & $0.021$ & $-0.792$ & $0.429$ & $-0.059$ & $-0.025$ \\
PhD Rank $+$ Bibliometric & $0.012$ & $0.021$ & $0.585$ & $0.559$ & $-0.029$ & $0.054$ \\
PhD Rank $+$ Bibliometric $+$ Co-authorship & $-0.009$ & $0.005$ & $-1.616$ & $0.106$ & $-0.019$ & $0.002$ \\
Co-authorship & $-0.124$ & $0.005$ & $-22.755$ & $0.000$ & $-0.134$ & $-0.113$ \\
Group Variance & $0.001$ & $0.015$ &  &  & & \\
\end{tabular}
}
\caption{\textbf{Linear mixed-effects modeling results for predicting faculty placement at top-$\mathbf{30}$ departments.} A summary of the linear mixed-effects regression model used to predict the PR-AUC of each model-feature combination as a function fo the feature set (fixed effect) and model architecture (random effect). We provide information relating to running the model including total number of observations and groups, the minimum, maximum, and mean group size, and the method. We consider four different reference feature sets (i.e., intercepts): (1) PhD rank, (2) bibliometric features, (3) PhD rank and bibliometric features, and (4) PhD rank and the co-authorhship network. Rows in blue highlight the feature sets which add the co-authorship network to the reference features and are used to generate Figure~\ref{fig:robustness}(a), while rows in green compare the PhD rank and bibliometric feature sets and are used to generate Figure~\ref{fig:robustness}(b).}
\label{tab:stats_y30}
\end{table}
%
% y_40 stats no rewiring
%
\begin{table}[t]
\centering
\resizebox{\textwidth}{!}{%
\begin{tabular}{lcccccc}
\multicolumn{7}{l}{\textbf{Mixed Linear Model Regression Summary}} \\
\\
\multicolumn{1}{l}{Model:} & \multicolumn{1}{l}{MixedLM} & \multicolumn{1}{l}{Dependent Variable:} & \multicolumn{1}{l}{PR-AUC} & \multicolumn{1}{l}{Method:} & \multicolumn{1}{l}{REML} & \\
\multicolumn{1}{l}{Scale:} & \multicolumn{1}{l}{0.0005} & \multicolumn{1}{l}{Log-Likelihood:} & \multicolumn{1}{l}{1104.9986} & \multicolumn{1}{l}{Converged:} & \multicolumn{1}{l}{Yes} & \\
\multicolumn{1}{l}{No. Observations:} & \multicolumn{1}{l}{480} & \multicolumn{1}{l}{No. Groups:} & \multicolumn{1}{l}{10} & \multicolumn{1}{l}{Min. group size:} & \multicolumn{1}{l}{20} & \\
\multicolumn{1}{l}{Max. group size:} & \multicolumn{1}{l}{80} & \multicolumn{1}{l}{Mean group size:} & \multicolumn{1}{l}{48.0} &  &  & \\
& & & & & & \\
\multicolumn{7}{l}{\textbf{Reference = PhD Rank}} \\
Term & Coef. & Std. Err. & $z$ & $P>|z|$ & CI Lower & CI Upper \\
\hline
Intercept & $0.806$ & $0.007$ & $110.209$ & $0.000$ & $0.792$ & $0.820$ \\
{\color[HTML]{009c4a}Bibliometric} & {\color[HTML]{009c4a}$-0.004$} & {\color[HTML]{009c4a}$0.005$} & {\color[HTML]{009c4a}$-0.881$} & {\color[HTML]{009c4a}$0.379$} & {\color[HTML]{009c4a}$-0.013$} & {\color[HTML]{009c4a}$0.005$} \\
Bibliometric $+$ Co-authorship & $0.014$ & $0.011$ & $1.242$ & $0.214$ & $-0.008$ & $0.035$ \\
PhD Rank $+$ Bibliometric & $0.019$ & $0.005$ & $4.108$ & $0.000$ & $0.010$ & $0.029$ \\
PhD Rank $+$ Bibliometric $+$ Co-authorship & $0.017$ & $0.011$ & $1.587$ & $0.113$ & $-0.004$ & $0.039$ \\
{\color[HTML]{3531FF}PhD Rank $+$ Co-authorship} & {\color[HTML]{3531FF}$0.011$} & {\color[HTML]{3531FF}$0.011$} & {\color[HTML]{3531FF}$1.007$} & {\color[HTML]{3531FF}$0.314$} & {\color[HTML]{3531FF}$-0.010$} & {\color[HTML]{3531FF}$0.032$} \\
Co-authorship & $-0.088$ & $0.011$ & $-8.115$ & $0.000$ & $-0.110$ & $-0.067$ \\
Group Variance & $0.000$ & $0.006$ &  &  & &  \\
& & & & & & \\
\multicolumn{7}{l}{\textbf{Reference = Bibliometric}} \\
Term & Coef. & Std. Err. & $z$ & $P>|z|$ & CI Lower & CI Upper \\
\hline
Intercept & $0.802$ & $0.007$ & $116.292$ & $0.000$ & $0.788$ & $0.815$ \\
{\color[HTML]{3531FF}Bibliometric $+$ Co-authorship} & {\color[HTML]{3531FF}$0.018$} & {\color[HTML]{3531FF}$0.011$} & {\color[HTML]{3531FF}$1.665$} & {\color[HTML]{3531FF}$0.096$} & {\color[HTML]{3531FF}$-0.003$} & {\color[HTML]{3531FF}$0.038$} \\
PhD Rank & $0.004$ & $0.005$ & $0.881$ & $0.379$ & $-0.005$ & $0.013$ \\
PhD Rank $+$ Bibliometric & $0.024$ & $0.004$ & $5.827$ & $0.000$ & $0.016$ & $0.031$ \\
PhD Rank $+$ Bibliometric $+$ Co-authorship & $0.021$ & $0.011$ & $2.019$ & $0.043$ & $0.001$ & $0.042$ \\
PhD Rank $+$ Co-authorship & $0.015$ & $0.011$ & $1.425$ & $0.154$ & $-0.006$ & $0.036$ \\
Co-authorship & $-0.084$ & $0.011$ & $-7.935$ & $0.000$ & $-0.105$ & $-0.063$ \\
Group Variance & $0.000$ & $0.006$ &  &  & &  \\
& & & & & & \\
\multicolumn{7}{l}{\textbf{Reference = PhD Rank $+$ Bibliometric}} \\
Term & Coef. & Std. Err. & $z$ & $P>|z|$ & CI Lower & CI Upper \\
\hline
Intercept & $0.825$ & $0.007$ & $119.703$ & $0.000$ & $0.812$ & $0.839$ \\
Bibliometric & $-0.024$ & $0.004$ & $-5.827$ & $0.000$ & $-0.031$ & $-0.016$ \\
Bibliometric $+$ Co-authorship & $-0.006$ & $0.011$ & $-0.550$ & $0.582$ & $-0.027$ & $0.015$ \\
PhD Rank & $-0.019$ & $0.005$ & $-4.108$ & $0.000$ & $-0.029$ & $-0.010$ \\
{\color[HTML]{3531FF}PhD Rank $+$ Bibliometric $+$ Co-authorship} & {\color[HTML]{3531FF}$-0.002$} & {\color[HTML]{3531FF}$0.011$} & {\color[HTML]{3531FF}$-0.196$} & {\color[HTML]{3531FF}$0.844$} & {\color[HTML]{3531FF}$-0.023$} & {\color[HTML]{3531FF}$0.019$} \\
PhD Rank $+$ Co-authorship & $-0.008$ & $0.011$ & $-0.791$ & $0.429$ & $-0.029$ & $0.012$ \\
Co-authorship & $-0.108$ & $0.011$ & $-10.151$ & $0.000$ & $-0.129$ & $-0.087$ \\
Group Variance & $0.000$ & $0.006$ &  &  & &  \\
& & & & & & \\
\multicolumn{7}{l}{\textbf{Reference = PhD Rank $+$ Co-authorship}} \\
Term & Coef. & Std. Err. & $z$ & $P>|z|$ & CI Lower & CI Upper \\
\hline
Intercept & $0.8817$ & $0.008$ & $101.175$ & $0.000$ & $0.801$ & $0.833$ \\
Bibliometric & $-0.015$ & $0.011$ & $-1.425$ & $0.154$ & $-0.036$ & $0.006$ \\
{\color[HTML]{009c4a}Bibliometric $+$ Co-authorship} & {\color[HTML]{009c4a}$0.003$} & {\color[HTML]{009c4a}$0.003$} & {\color[HTML]{009c4a}$0.731$} & {\color[HTML]{009c4a}$0.465$} & {\color[HTML]{009c4a}$-0.004$} & {\color[HTML]{009c4a}$0.009$} \\
PhD Rank & $-0.011$ & $0.011$ & $-1.007$ & $0.314$ & $-0.032$ & $0.010$ \\
PhD Rank $+$ Bibliometric & $0.008$ & $0.011$ & $0.791$ & $0.429$ & $-0.012$ & $0.029$ \\
PhD Rank $+$ Bibliometric $+$ Co-authorship & $0.006$ & $0.003$ & $1.806$ & $0.071$ & $-0.001$ & $0.013$ \\
Co-authorship & $-0.099$ & $0.003$ & $-28.427$ & $0.000$ & $-0.106$ & $-0.093$ \\
Group Variance & $0.000$ & $0.006$ &  &  & &  \\
\end{tabular}
}
\caption{\textbf{Linear mixed-effects modeling results for predicting faculty placement at top-$\mathbf{40}$ departments.} A summary of the linear mixed-effects regression model used to predict the PR-AUC of each model-feature combination as a function fo the feature set (fixed effect) and model architecture (random effect). We provide information relating to running the model including total number of observations and groups, the minimum, maximum, and mean group size, and the method. We consider four different reference feature sets (i.e., intercepts): (1) PhD rank, (2) bibliometric features, (3) PhD rank and bibliometric features, and (4) PhD rank and the co-authorhship network. Rows in blue highlight the feature sets which add the co-authorship network to the reference features and are used to generate Figure~\ref{fig:robustness}(a), while rows in green compare the PhD rank and bibliometric feature sets and are used to generate Figure~\ref{fig:robustness}(b).}
\label{tab:stats_y40}
\end{table}
%
% y_50 stats no rewiring
%
\begin{table}[t]
\centering
\resizebox{\textwidth}{!}{%
\begin{tabular}{lcccccc}
\multicolumn{7}{l}{\textbf{Mixed Linear Model Regression Summary}} \\
\\
\multicolumn{1}{l}{Model:} & \multicolumn{1}{l}{MixedLM} & \multicolumn{1}{l}{Dependent Variable:} & \multicolumn{1}{l}{PR-AUC} & \multicolumn{1}{l}{Method:} & \multicolumn{1}{l}{REML} & \\
\multicolumn{1}{l}{Scale:} & \multicolumn{1}{l}{0.0003} & \multicolumn{1}{l}{Log-Likelihood:} & \multicolumn{1}{l}{1255.6296} & \multicolumn{1}{l}{Converged:} & \multicolumn{1}{l}{Yes} & \\
\multicolumn{1}{l}{No. Observations:} & \multicolumn{1}{l}{480} & \multicolumn{1}{l}{No. Groups:} & \multicolumn{1}{l}{10} & \multicolumn{1}{l}{Min. group size:} & \multicolumn{1}{l}{20} & \\
\multicolumn{1}{l}{Max. group size:} & \multicolumn{1}{l}{80} & \multicolumn{1}{l}{Mean group size:} & \multicolumn{1}{l}{48.0} &  &  & \\
& & & & & & \\
\multicolumn{7}{l}{\textbf{Reference = PhD Rank}} \\
Term & Coef. & Std. Err. & $z$ & $P>|z|$ & CI Lower & CI Upper \\
\hline
Intercept & $0.878$ & $0.007$ & $125.080$ & $0.000$ & $0.864$ & $0.892$ \\
{\color[HTML]{009c4a}Bibliometric} & {\color[HTML]{009c4a}$-0.001$} & {\color[HTML]{009c4a}$0.003$} & {\color[HTML]{009c4a}$-0.230$} & {\color[HTML]{009c4a}$0.818$} & {\color[HTML]{009c4a}$-0.007$} & {\color[HTML]{009c4a}$0.006$} \\
Bibliometric $+$ Co-authorship & $-0.004$ & $0.011$ & $-0.326$ & $0.744$ & $-0.025$ & $0.018$ \\
PhD Rank $+$ Bibliometric & $0.011$ & $0.003$ & $3.365$ & $0.001$ & $0.005$ & $0.018$ \\
PhD Rank $+$ Bibliometric $+$ Co-authorship & $0.013$ & $0.011$ & $1.223$ & $0.221$ & $-0.008$ & $0.034$ \\
{\color[HTML]{3531FF}PhD Rank $+$ Co-authorship} & {\color[HTML]{3531FF}$0.017$} & {\color[HTML]{3531FF}$0.011$} & {\color[HTML]{3531FF}$1.567$} & {\color[HTML]{3531FF}$0.117$} & {\color[HTML]{3531FF}$-0.004$} & {\color[HTML]{3531FF}$0.038$} \\
Co-authorship & $-0.048$ & $0.011$ & $-4.509$ & $0.000$ & $-0.069$ & $-0.027$ \\
Group Variance & $0.000$ & $0.008$ &  & & &  \\
& & & & & & \\
\multicolumn{7}{l}{\textbf{Reference = Bibliometric}} \\
Term & Coef. & Std. Err. & $z$ & $P>|z|$ & CI Lower & CI Upper \\
\hline
Intercept & $0.877$ & $0.007$ & $129.160$ & $0.000$ & $0.864$ & $0.890$ \\
{\color[HTML]{3531FF}Bibliometric $+$ Co-authorship} & {\color[HTML]{3531FF}$-0.003$} & {\color[HTML]{3531FF}$0.011$} & {\color[HTML]{3531FF}$-0.257$} & {\color[HTML]{3531FF}$0.797$} & {\color[HTML]{3531FF}$-0.023$} & {\color[HTML]{3531FF}$0.018$} \\
PhD Rank & $0.001$ & $0.003$ & $0.230$ & $0.818$ & $-0.006$ & $0.007$ \\
PhD Rank $+$ Bibliometric & $0.012$ & $0.003$ & $4.205$ & $0.000$ & $0.007$ & $0.018$ \\
PhD Rank $+$ Bibliometric $+$ Co-authorship & $0.014$ & $0.011$ & $1.314$ & $0.189$ & $-0.007$ & $0.035$ \\
PhD Rank $+$ Co-authorship & $0.018$ & $0.011$ & $1.663$ & $0.096$ & $-0.003$ & $0.038$ \\
Co-authorship & $-0.048$ & $0.011$ & $-4.498$ & $0.000$ & $-0.068$ & $-0.027$ \\
Group Variance & $0.000$ & $0.008$ &  & & &  \\
& & & & & & \\
\multicolumn{7}{l}{\textbf{Reference = PhD Rank $+$ Bibliometric}} \\
Term & Coef. & Std. Err. & $z$ & $P>|z|$ & CI Lower & CI Upper \\
\hline
Intercept & $0.889$ & $0.007$ & $130.968$ & $0.000$ & $0.876$ & $0.903$ \\
Bibliometric & $-0.012$ & $0.003$ & $-4.205$ & $0.000$ & $-0.018$ & $-0.007$ \\
Bibliometric $+$ Co-authorship & $-0.015$ & $0.011$ & $-1.417$ & $0.157$ & $-0.036$ & $0.006$ \\
PhD Rank & $-0.011$ & $0.003$ & $-3.365$ & $0.001$ & $-0.018$ & $-0.005$ \\
{\color[HTML]{3531FF}PhD Rank $+$ Bibliometric $+$ Co-authorship} & {\color[HTML]{3531FF}$0.002$} & {\color[HTML]{3531FF}$0.011$} & {\color[HTML]{3531FF}$0.154$} & {\color[HTML]{3531FF}$0.877$} & {\color[HTML]{3531FF}$-0.019$} & {\color[HTML]{3531FF}$0.022$} \\
PhD Rank $+$ Co-authorship & $0.005$ & $0.011$ & $0.503$ & $0.615$ & $-0.015$ & $0.026$ \\
Co-authorship & $-0.060$ & $0.011$ & $-5.657$ & $0.000$ & $-0.081$ & $-0.039$ \\
Group Variance & $0.000$ & $0.008$ &  & & &  \\
& & & & & & \\
\multicolumn{7}{l}{\textbf{Reference = PhD Rank $+$ Co-authorship}} \\
Term & Coef. & Std. Err. & $z$ & $P>|z|$ & CI Lower & CI Upper \\
\hline
Intercept & $0.895$ & $0.008$ & $110.151$ & $0.000$ & $0.879$ & $0.911$ \\
Bibliometric & $-0.018$ & $0.011$ & $-1.663$ & $0.096$ & $-0.038$ & $0.003$ \\
{\color[HTML]{009c4a}Bibliometric $+$ Co-authorship} & {\color[HTML]{009c4a}$-0.020$} & {\color[HTML]{009c4a}$0.003$} & {\color[HTML]{009c4a}$-8.038$} & {\color[HTML]{009c4a}$0.000$} & {\color[HTML]{009c4a}$-0.025$} & {\color[HTML]{009c4a}$-0.015$} \\
PhD Rank & $-0.017$ & $0.011$ & $-1.567$ & $0.117$ & $-0.038$ & $-0.004$ \\
PhD Rank $+$ Bibliometric & $-0.005$ & $0.011$ & $-0.503$ & $0.615$ & $-0.026$ & $0.015$ \\
PhD Rank $+$ Bibliometric $+$ Co-authorship & $-0.004$ & $0.003$ & $-1.461$ & $0.144$ & $-0.009$ & $0.001$ \\
Co-author & $-0.065$ & $0.003$ & $-25.794$ & $0.000$ & $-0.070$ & $-0.060$ \\
Group Variance & $0.000$ & $0.008$ &  & & &  \\
\end{tabular}
}
\caption{\textbf{Linear mixed-effects modeling results for predicting faculty placement at top-$\mathbf{50}$ departments.} A summary of the linear mixed-effects regression model used to predict the PR-AUC of each model-feature combination as a function fo the feature set (fixed effect) and model architecture (random effect). We provide information relating to running the model including total number of observations and groups, the minimum, maximum, and mean group size, and the method. We consider four different reference feature sets (i.e., intercepts): (1) PhD rank, (2) bibliometric features, (3) PhD rank and bibliometric features, and (4) PhD rank and the co-authorhship network. Rows in blue highlight the feature sets which add the co-authorship network to the reference features and are used to generate Figure~\ref{fig:robustness}(a), while rows in green compare the PhD rank and bibliometric feature sets and are used to generate Figure~\ref{fig:robustness}(b).}
\label{tab:stats_y50}
\end{table}

\end{appendices}

\end{document}